\documentclass[12pt,preprint]{aastex}
\shorttitle{Off-limb formation of the He {\sc i} 10830 \AA\ and D$_3$ multiplets}
\shortauthors{Centeno et al.}

\begin{document}

\title{The influence of coronal EUV irradiance on the emission in the He {\sc i} 10830 \AA\ and D$_3$ multiplets}

\author{R. Centeno\altaffilmark{1,2}, J. Trujillo Bueno\altaffilmark{1,3}, H. Uitenbroek\altaffilmark{4} and M. Collados\altaffilmark{1}}

\altaffiltext{1}{Instituto de Astrof\'{\i}sica de Canarias, 38205 La Laguna, Tenerife, Spain}
\altaffiltext{2}{High Altitude Observatory (NCAR), Boulder CO 80301, USA}
\altaffiltext{3}{Consejo Superior de Investigaciones Cient\'{\i}ficas (Spain)}
\altaffiltext{4}{National Solar Observatory, Sac. Peak, NM, USA}
\email{rce@ucar.edu, jtb@iac.es, huitenbr@nso.edu, mcv@iac.es}

\begin{abstract}
Two of the most attractive spectral windows for spectropolarimetric 
investigations of the physical properties of the plasma structures in the 
solar chromosphere and corona are the ones provided by the spectral 
lines of the He {\sc i} 10830 \AA\ and 5876 \AA\ (or D$_3$) multiplets, 
whose polarization signals are sensitive to the Hanle and Zeeman effects.
However, in order to be able to carry out reliable diagnostics, it is crucial 
to have a good physical understanding of the sensitivity of the observed 
spectral line radiation to the various competing driving mechanisms.
Here we report a series of off-the-limb non-LTE 
calculations of the He {\sc i} D$_3$ and  10830 \AA\ emission profiles, 
focusing our investigation on their sensitivity to the EUV coronal 
irradiation and the model atmosphere used in the calculations. 
We show in particular that the intensity ratio of the blue to the red 
components in
the emission profiles of the He {\sc i} 10830 \AA\ multiplet turns out to be 
a good candidate as a diagnostic 
tool for the coronal irradiance. Measurements of this observable as a function 
of the distance to the limb and its confrontation with radiative transfer 
modeling might give us valuable information on the physical properties of 
the solar atmosphere and on the amount of EUV radiation at relevant wavelengths
penetrating the chromosphere from above.

\end{abstract}


\section{Introduction}

The spectral lines of the multiplet of neutral helium at 10830 \AA\ are of 
great diagnostic value for investigating the dynamic and magnetic properties 
of plasma structures in the solar chromosphere and corona. This is because 
their polarization is sensitive to the presence of atomic level polarization 
and to the joint action of the Hanle and Zeeman effects
(e.g. Trujillo Bueno \& Asensio Ramos, 2007), which makes these lines 
especially sensitive to the strongest fields found in solar active regions 
(e.g., Harvey \& Hall 1977; R\"uedi et al. 1995; Centeno et al. 2006) and 
also to the weaker fields encountered in a variety of plasma 
structures such as filaments (e.g., Lin et al. 1998; Trujillo 
Bueno et al. 2002), regions of emerging magnetic flux (e.g., Lagg et al. 
2004), chromospheric spicules (e.g., Trujillo Bueno et al. 2005; 
Socas-Navarro et al. 2005) 
and prominences (e.g., Merenda et al. 2006). Obviously, in order to be able 
to obtain reliable inferences from the observed spectral line radiation it 
is necessary to reach a rigorous physical understanding of the key 
mechanisms that are responsible for its intensity and polarization. The 
same applies to the lines of the He {\sc i} D$_3$ multiplet at 5876 \AA\, 
whose observed polarization has been used for investigating 
the magnetic field vector in solar prominences and spicules (e.g., Landi 
Degl'Innocenti 1982; Leroy, Bommier \& Sahal-Br\'echot, 1983; Casini et al. 
2003; L\'opez Ariste \& Casini 2005; Bianda et al. 2006; Ramelli et al. 
2006 a,b).

While the modeling of the spectral line intensity has been done by solving 
the standard non-LTE radiation transfer problem in one-dimensional (1-D)
semi-empirical models of the solar atmosphere (e.g., Pozhalova 1988; 
Avrett et al. 1994; 
Andretta \& Jones 1997), the physical interpretation of the observed spectral 
line polarization has been carried out taking into account the joint 
action of the Hanle and Zeeman effects within the framework of the quantum
theory of polarization, but  simplifying the radiative 
transfer problem by using the optically thin approximation (e.g., Landi
Degl'Innocenti, 1982; Leroy et al, 1983;
Casini et al 2004) or by solving the Stokes-vector transfer equation in
a slab of constant physical properties (Trujillo Bueno et al. 2002; 2005). 

In this slab model for the interpretation of polarization 
observations, the slab's optical thickness is simply chosen as a free 
parameter to fit the observed intensity profile. This ``cloud" model 
for the interpretation of polarization observations simplifies 
the forward modeling and inversion problem (see Trujillo Bueno \& Asensio 
Ramos, 2007; Asensio Ramos \& Trujillo 
Bueno 2007), but it does not give us 
information on the possible errors we make by not considering explicitly the 
physical mechanisms that are thought to be responsible for producing some 
optical thickness in such lines of neutral helium. 

\noindent According to the 
above-mentioned non-LTE radiative transfer investigations in 1D 
semi-empirical models of the solar atmosphere there are two candidates 
for producing a significant optical thickness in the lines of the 
10830 \AA\ and D$_3$ multiplets, given that the collisional excitations 
at typical solar chromospheric temperatures ($<$20000 K) are unable to 
significantly populate the triplet levels of He {\sc i}. The first 
mechanism states that the EUV radiation irradiating the chromosphere 
ionizes part of the substantial amount of neutral helium 
atoms that are found in the ground singlet state at the typical chromospheric 
temperatures, which then lead to recombinations that produce an enhanced 
population of the triplet states (e.g., Pozhalova 1988; Avrett et al. 1994). 
The second one 
deals with the possibility of collisional excitation by a significant amount of 
transition region material at temperatures greater than 20000 K (e.g., 
Andretta \& Jones 1997). In our opinion, as we shall try to show 
in this paper, detailed off-limb observations of the intensity and 
polarization profiles of the He {\sc i} 10830 \AA\ multiplet should 
help us determine which is the dominating mechanism, or if both 
are playing an equally significant role. 

The mechanism of collisional excitation at temperatures greater than 20000 K 
seems to require (or is helped by) the existence of a temperature ``plateau" 
located in the lower transition region (TR) of the Sun, such as that included 
ad-hoc by Vernazza et al. (1981) in their semi-empirical models in order to 
be able to model the observed Ly$_{\alpha}$ radiation. Interestingly enough, 
the consideration of ambipolar diffusion effecs in such semi-empirical 
models led to a new series of models by Fontenla et al. (1990; 1991; 1993) 
that are able to explain the observed Ly$_{\alpha}$ intensity observations 
without the need of such a TR ``plateau". In this new series of semi-empirical 
models (FAL models), an enhanced population in the triplet states 
of He {\sc i} can only be produced by the above mentioned 
Photoionization-Recombination process (hereafter: the PR mechanism). 

The main aim of this paper is to investigate the sensitivity of the 
off-the-limb emission profiles of the 10830 \AA\ and D$_3$ lines to the 
amount of coronal EUV radiation. To this end, we have chosen the 
above-mentioned FAL-like models and have solved the RT problem in 
spherical geometry for increasing values of the EUV coronal irradiance. 
As we shall see, this has allowed us to demonstrate that the intensity 
ratio of the blue and red components of the 10830 \AA\ multiplet is a 
sensitive function of the amount of coronal EUV illumination, and to 
establish this very observable as an ideal candidate for mapping 
the amount of coronal EUV irradiance at more or less active 
off-limb locations and for determining whether or not the PR mechanism 
is truly the dominating one. We also show the variation 
of this intensity ratio with the atmospheric height when using an atmospheric 
model with a TR ``plateau" in the model's temperature profile. In our 
opinion, the comparison of this type of simulations with observations 
should allow us to determine whether there is any off-limb region of the 
Sun (e.g., above plages?) where collisional excitation plays a 
significant role in addition to the PR mechanism.

The outline of this paper is as follows. The formulation of the problem 
is presented in Sect. 2, where we describe our approach to model the off-limb emission profiles 
of the aforementioned multiplets. Section 3 focuses on the sensitivity
of these profiles to the EUV coronal irradiation and points out the 
existence of an interesting observable (i.e., the above-mentioned
intensity ratio) that can serve as diagnostic tool
for this physical quantity. Finally, in section 4 we summarize the highlights
and discuss possible lines of future research.

\section{Formulation of the problem}

For understanding the formation of these helium lines
we carried out a series of non-LTE radiative transfer calculations of the 
off-limb intensity profiles emerging from various semi-empirical 
atmospheric models using a 1D spherically symmetric geometry version of 
the RH code (see {\em http://www.nso.edu/staff/uitenbr/rh.html}). 
This non-LTE radiative transfer code, based on the Multi-level Accelerated
Lambda Iteration (MALI) scheme by Rybicki \& Hummer (1991, 1992), solves
the combined equations of statistical equilibrium and radiative transfer
for a multilevel atom in a given stellar atmospheric model
(see also Socas-Navarro and Trujillo Bueno, 1997). The code
deals with overlapping transitions, and besides from the opacities and
emissivities produced by the transitions in the {\em active} atom 
(the one that is considered in full non-LTE regime)
it also accounts for background radiation due to other
atoms, molecules and relevant continuum processes.

\begin{figure}[!t]
\begin{center}
\includegraphics[scale=.50]{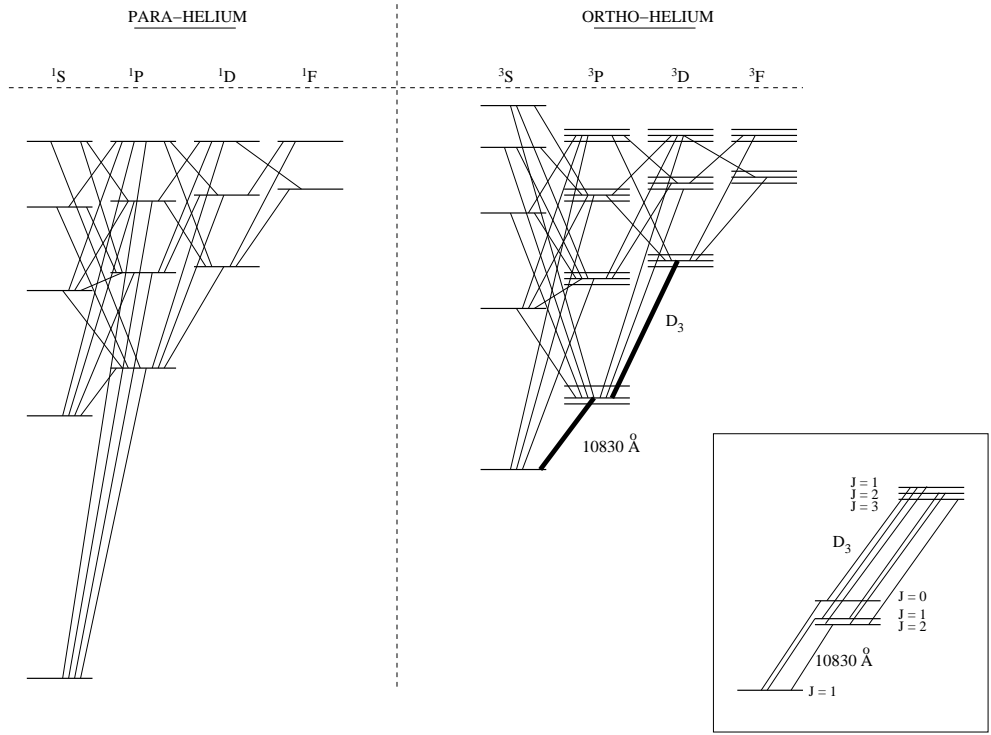}
\caption{Schematic atomic model of He {\sc i}, with 46 energy levels and
the transitions between each pair of levels. Although
the multiple transitions among triplet levels are not represented
in the figure, they were taken into account in the calculations. 
The individual He {\sc i} 
10830 \AA\ and D$_3$ transitions are highlighted in the inset.}
\label{fig:atomic-model}
\end{center}
\end{figure}

The calculations of the emergent emission spectra were done in the 
context of the FAL model atmospheres (Fontenla et al, 1990,
1991, 1993), which require the above-mentioned PR mechanism for producing
significant off-limb emission in the He {\sc i} 10830 \AA\ and $D_3$ 
multiplets. We have used these semi-empirical models as tabulated (i.e.
without considering any possible feedback effect of the EUV irradiance
on the models themselves). The main analysis is carried out with the 
FAL-C model of the 
average quiet Sun, although a comparison of the results among  
FAL-C, FAL-P (representative of a plage region) and the M$_{\rm CO}$ model
of Avrett (i.e., the relatively cool M$_{\rm CO}$ model atmosphere illustrative of 
the quiet Sun that Avrett (1995) proposed to account for the 
observed molecular CO absorption at 4.6 $\mu$m) is described at the end 
of section 3. In addition, we show also a comparison with the results
obtained in the VAL-C model of Vernazza et al. (1981), whose temperature
plateau (with $T>20000$ K) in its TR to the coronal gas, gives rise to a 
localized collisional enhancement of the triplet states populations.

\noindent The atomic model of He that we used 
(see Fig. \ref{fig:atomic-model}) accounts for 165 possible
bound-bound transitions taking place among 53 energy levels (46 of 
He {\sc i}, 6 of He {\sc ii} and 1 of He {\sc iii}). The fine structure
of the triplet energy levels was taken into account, together with 
all the possible combinations of transitions between them.
The radiative transition parameters were obtained from the 
NIST database\footnote{National Institute for Standards and Technology, see
{\em http://physics.nist.gov/PhysRefData/ASD/index.html}} and 
the collisional rate coefficients were calculated using the Seaton (1962) 
impact approximation for neutral bound-bound and the Van Regemorter's (1962) 
approximation for treating bound-bound transitions caused by collisions
with electrons.
Both photoionization and collisional ionization bound-free cross-sections are
also tabulated in the atomic model file and were extracted from 
Landolt-B\"ornstein (1982).

\begin{figure}[!t]
\begin{center}
\includegraphics[scale=.60]{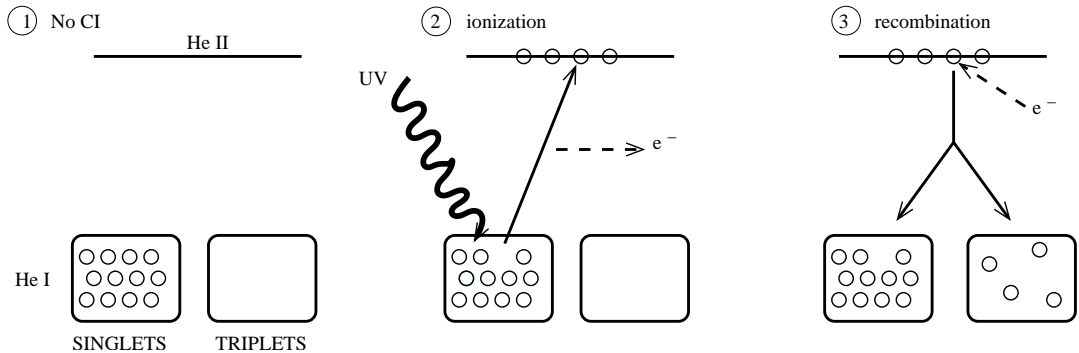}
\caption{Ionization-recombination scheme for the He atom triggered by the EUV coronal irradiance (CI). In the absence of CI (panel 1) nearly all the population of He is in the ground state of the singlet system. The photoionization-recombination process (panels 2 and 3) is able to populate the triplet system significantly.}\label{fig:recombination}
\end{center}
\end{figure}

Despite its simplicity from the atomic point of view, the atom of Helium
shows quite a complicated spectrum with two different spectral series: 
one of singlets (Para-Helium) involving singlet terms ($^1S$, 
$^1P$, $^1D$..) and the other one of triplets (Ortho-Helium) involving
triplet terms ($^3S$, $^3P$, $^3D$..).
In the electric dipole approximation, quantum selection rules do not
allow radiative transitions between singlet and triplet terms, so the
populations of the two sets of levels are not radiatively coupled. For this
reason, the lower energy level in the triplet system is a metastable one.
At typical solar chromospheric conditions, the bulk of the Helium population is
mainly concentrated in the $^1$S atomic ground level since the temperature is
not high enough for populating the rest of the levels significantly.
Thus, the insufficent population of the triplet system cannot account for the 
observed He {\sc i} 10830 \AA\ and 5876 \AA\ features.
It has been pointed out that the properties of the He {\sc i} 10830 \AA\ and 
$D_3$ multiplets depend mainly on the density and thickness of the 
chromosphere as well as on the incoming 
EUV irradiation from the corona that incides on the chromosphere 
(see Pozhalova 1988; Avrett et al. 1994; and earlier references therein). 
This EUV spectrum with $\lambda < 504$ \AA\ ionizes 
Helium increasing the He {\sc ii} number density, which subsequently recombine
with the free electrons populating in this way both singlet and triplet systems
(Fig. \ref{fig:recombination} describes this process in 3 schematic steps).
The excess population in the triplet energy levels is responsible for 
strengthening the He {\sc i} 10830 \AA\ and $D_3$ features.

\begin{figure}[!t]
\begin{center}
\includegraphics[scale=.50]{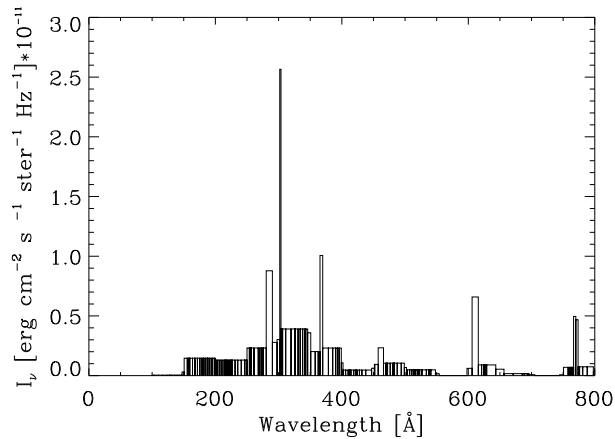}
\caption{EUV flux incident on the chromosphere based on Tobiska (1991).}
\label{fig:tobiska}
\end{center}
\end{figure}

The inclusion of coronal irradiance incident on the chromosphere 
was implemented in the RH code using the EUV flux given
by Tobiska (1991). This author compiled data -provided by six different 
satellites-
of the EUV coronal irradiance that reached the Earth in the period
between 1962 and 1991. 
These data were used to evaluate the chromospheric illumination due to coronal
sources, which is shown in Fig. \ref{fig:tobiska}.

\section{Results of the RT modeling calculations}

Following the strategy described in the last section
we carried out a thorough analysis of the sensitivity of the He {\sc i}
10830 \AA\ and D$_3$ emission profiles to the ionizing coronal irradiance.
We will show, in particular, how off-limb observations of the He {\sc i}
10830 \AA\ multiplet can serve as a diagnostic tool for this EUV 
irradiance and discuss 
their use as a constraint for future atmospheric models.

\subsection{He {\sc i} 10830 \AA\ and D3 emission profiles}

In a first approach to understand the off-limb emission of the lines of the 
10830 \AA\ and D$_3$ multiplets we compute the emergent profiles from the 
FAL-C model atmosphere.
In order to calculate the non-LTE populations of He {\sc i} and the emergent 
profiles we used a version of the RH code in 1D spherically symmetric geometry.
We solved consistently the RT in spherical geometry
and the SE equations to determine the populations at each height, and
computed the emergent intensity profiles on a set of rays tangent to
each spherical shell of the model atmosphere.

\begin{figure}[t!]
\centering
\includegraphics[width=7cm]{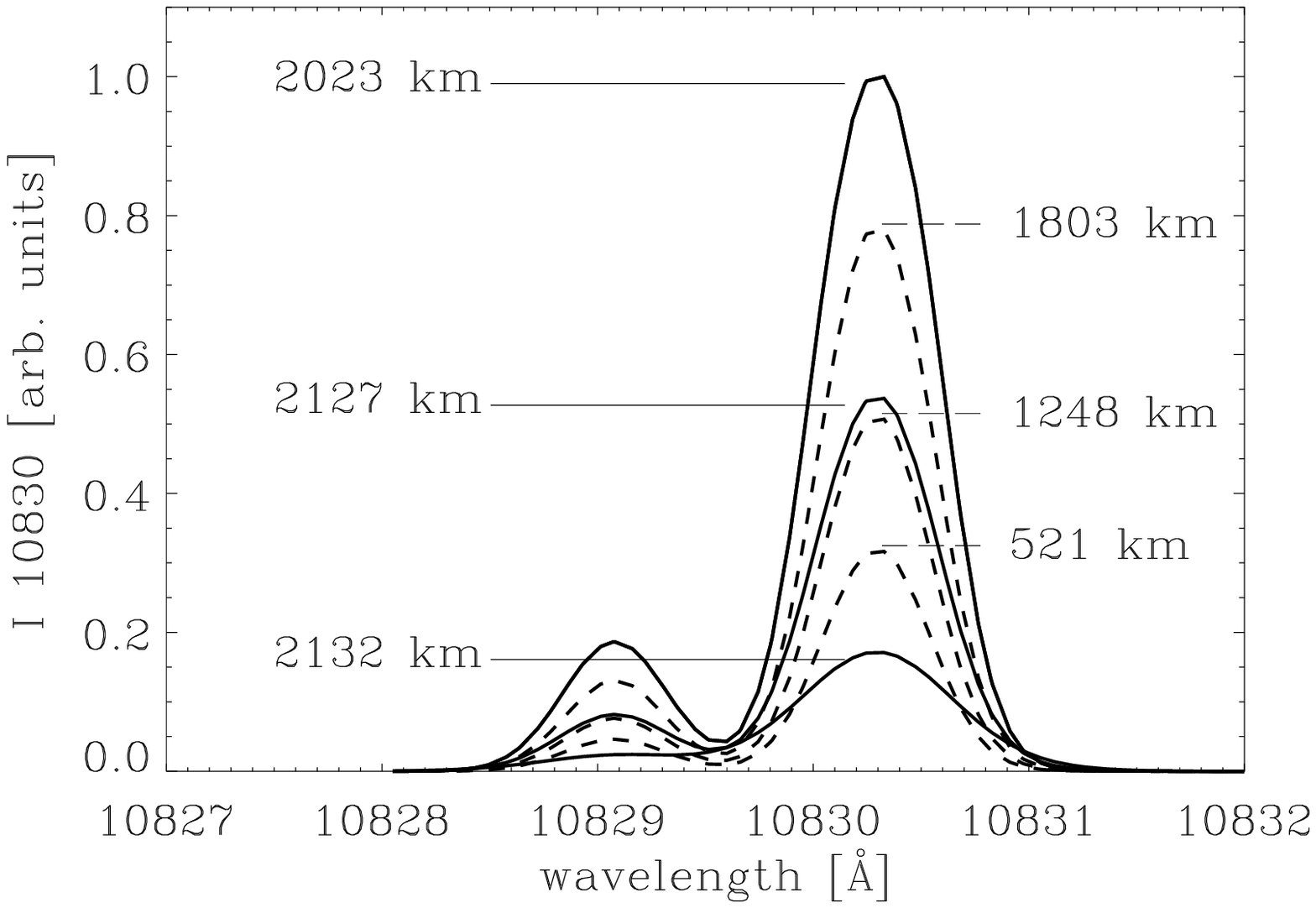}
\includegraphics[width=7cm]{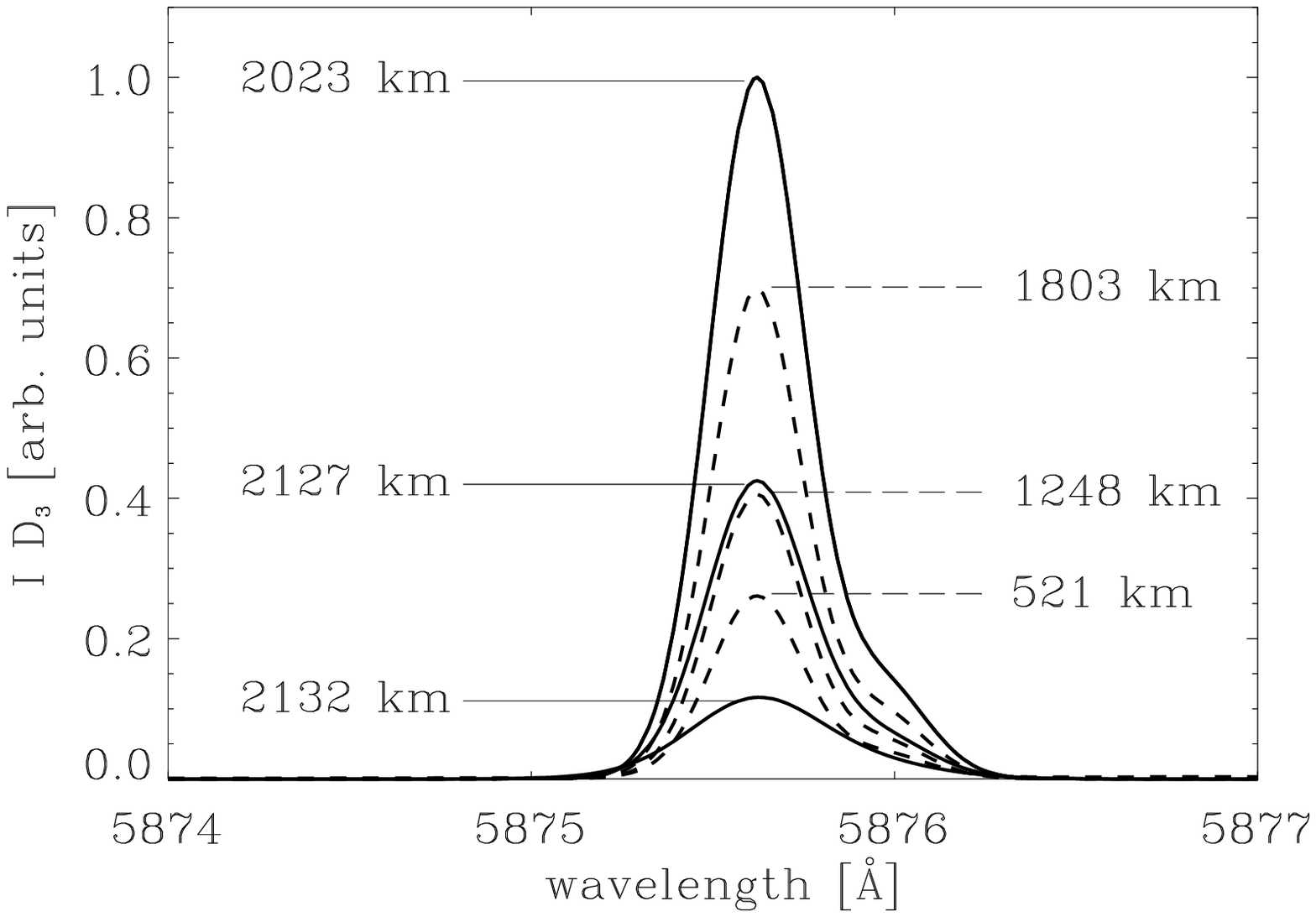}
\protect\caption[ ]{Off-the-limb He {\sc i} 10830 \AA\ (left) and $D_3$ 
5876 \AA\ (right) emission profiles at different distances to the limb
of the FAL-C semi-empirical model, when the nominal coronal irradiance
has been applied to the system. 
Heights are given relative to the reference $\tau_{500} = 1$. 
\label{fig:offlimb}}
\end{figure}

Fig. \ref{fig:offlimb} represents the He {\sc i} 10830 \AA\ (left) and
5876 \AA\ (right) profiles with increasing distance to the base of the 
photosphere\footnote{The height reference
is set to be the level at which the optical depth -in the vertical direction- 
of the continuum at 
500 nm equals unity, $\tau_{500}=1$} when the incident coronal
irradiation takes its nominal value (see Fig. \ref{fig:tobiska}). 
The lower height
corresponds to the first profile seen in emission off the limb.
In the outermost layers of the FAL-C model atmosphere, both multiplet
transitions produce very little emission. As the distance to the limb 
decreases, the amplitude of the emergent profiles grows until it reaches
a maximum around 2000 km. Below this point, the emission
starts decreasing slowly until it turns into absorption inside the limb.
The interpretation is as follows: the density in the outer layers of the 
atmosphere (at large distances from the limb) is so low that it cannot 
produce measurable emission profiles. As we go deeper into the atmosphere,
the density increases quickly and so does the emission in the multiplets,
until it reaches a maximum at a height of $\sim 2000$ km. The extinction
of the EUV radiation as it travels inwards through the chromosphere 
leads to a reduction in the number of ionizations in the inner layers. Thus,
the emission in the spectral lines of the He {\sc i} 10830 \AA\ and D$_3$ 
multiplets starts
decreasing again because the PR mechanism cannot sustain the populations 
of the triplet system.

\subsection{Dependence with coronal irradiance}

One of the purposes of this investigation is to study the sensitivity
of the emission profiles to the amount of coronal irradiance (CI).

\begin{figure}[t!]
\centering
\includegraphics[width=7cm]{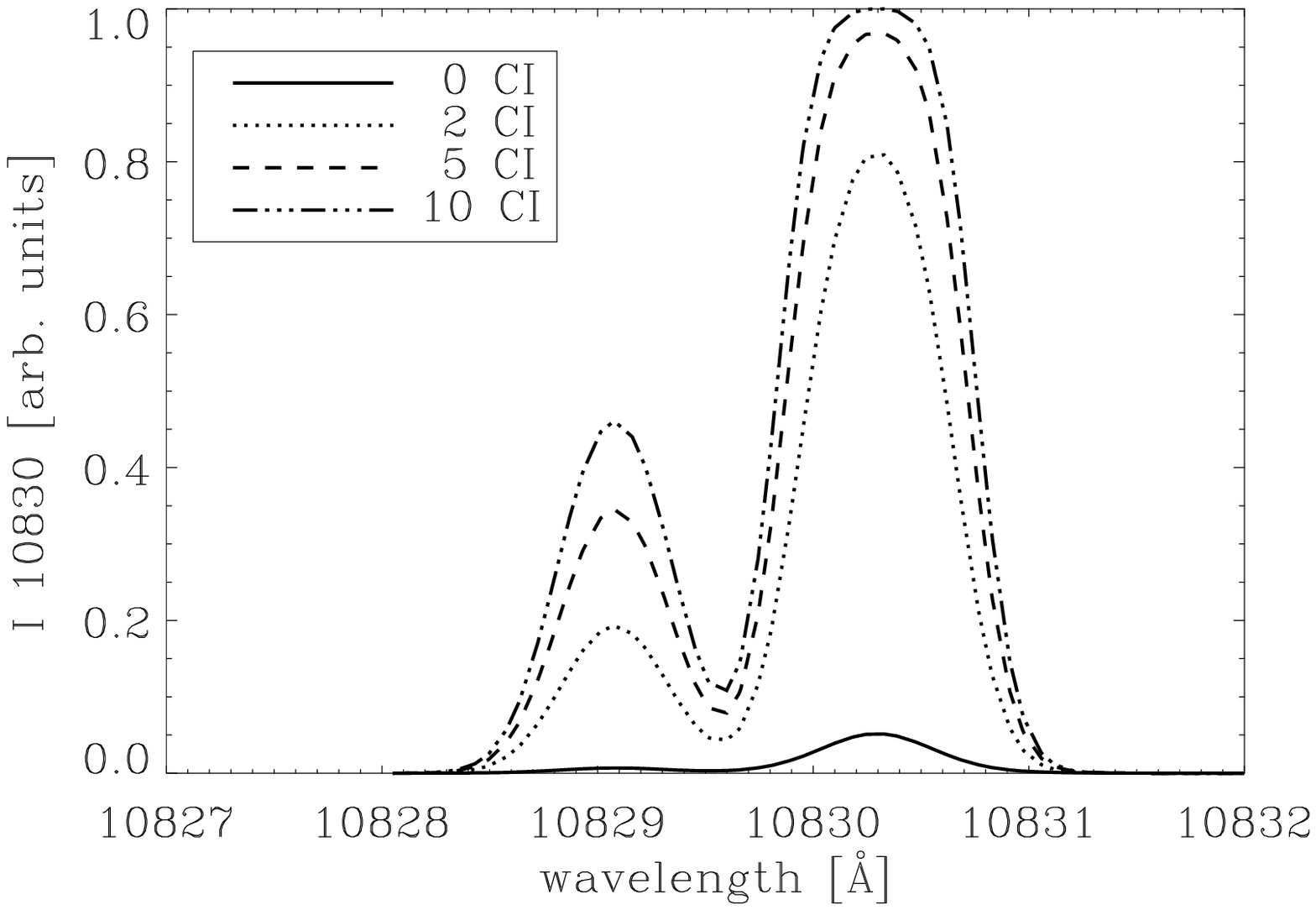}
\includegraphics[width=7cm]{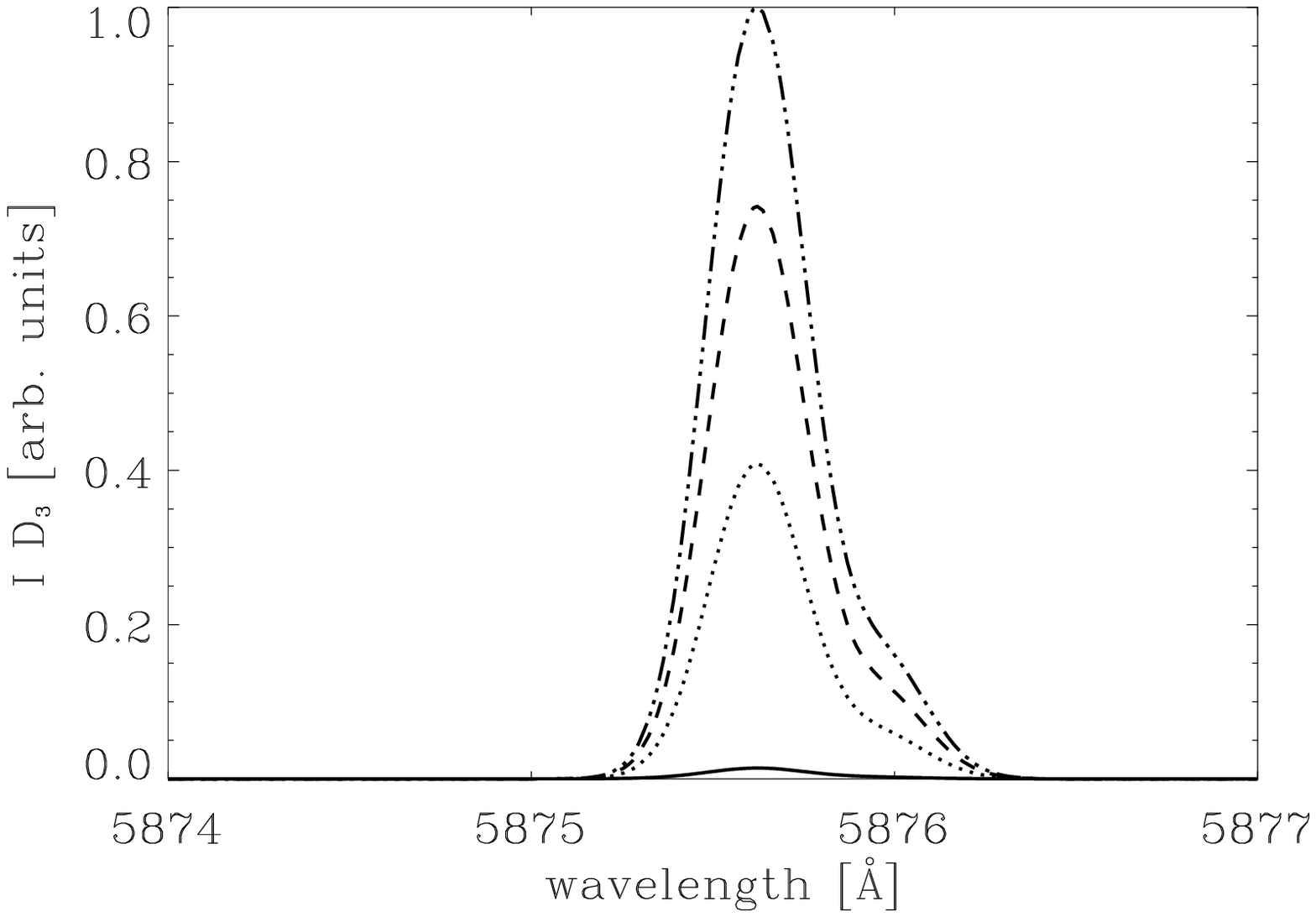}\\
\includegraphics[width=7cm]{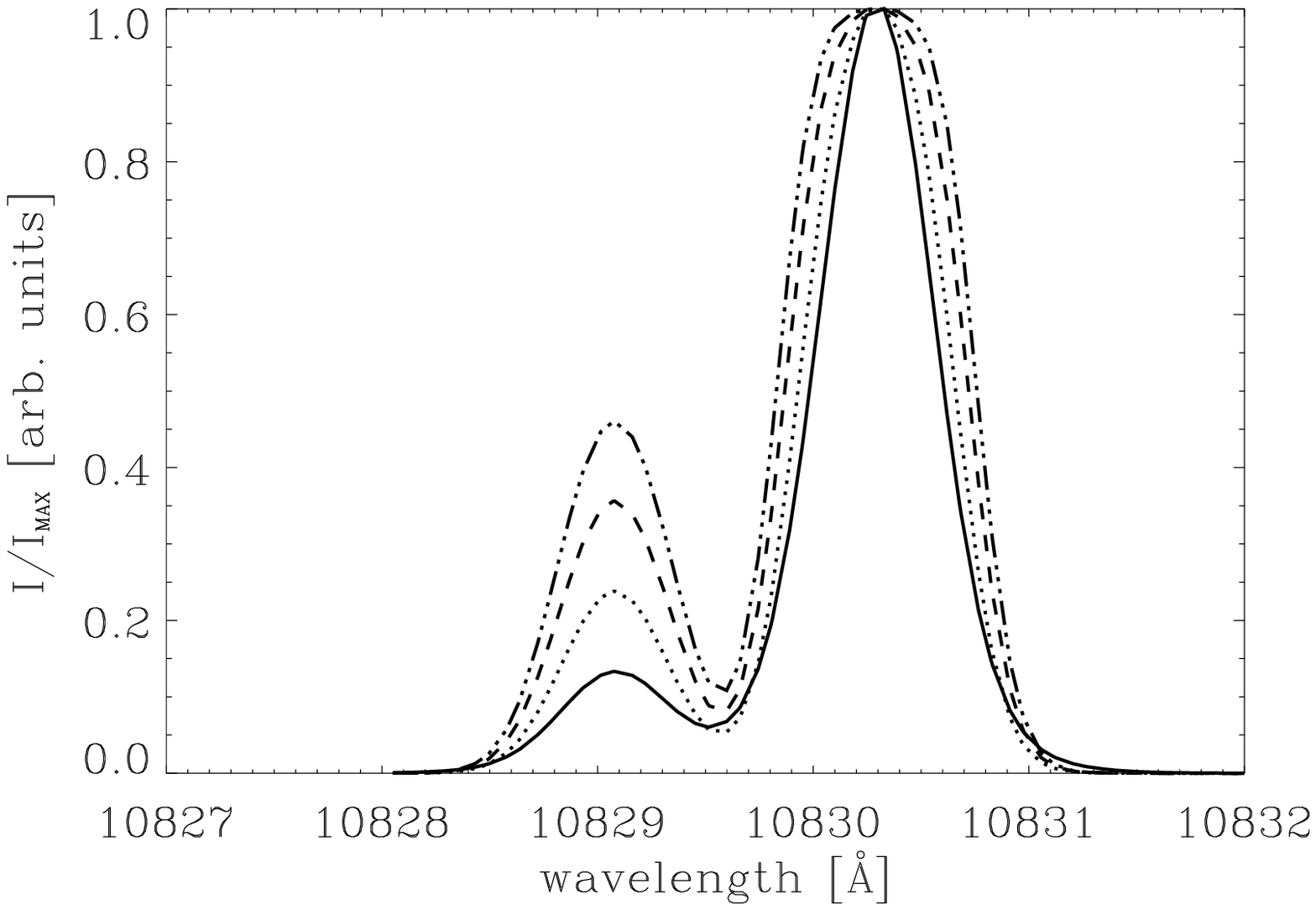}
\includegraphics[width=7cm]{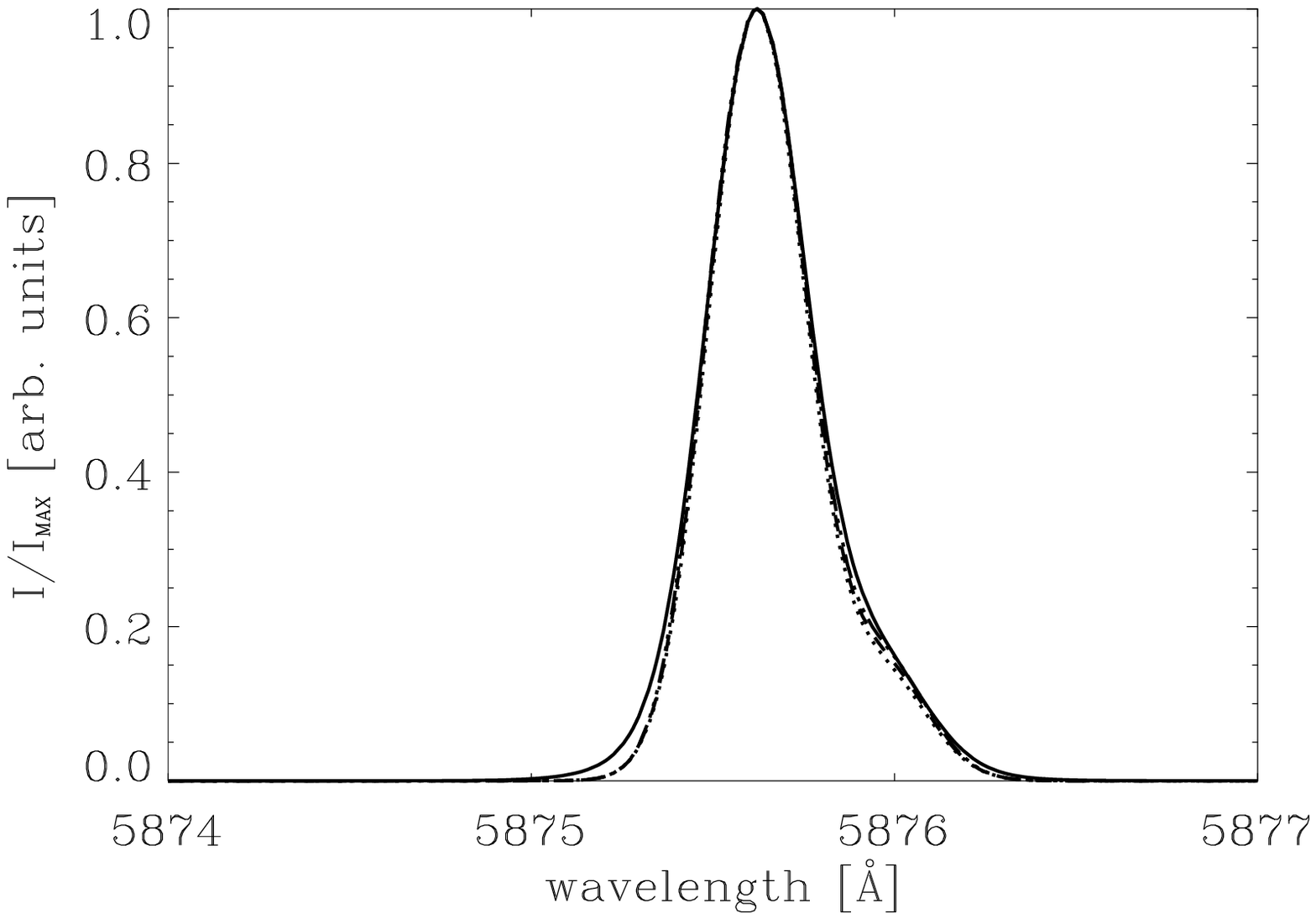}
\protect\caption[ ]{{\em Top:} Emission intensity profiles of He {\sc i} 
10830 \AA\ ({\em left}) and $D_3$ ({\em right}) at a height of 2023 km above
the $\tau_{500}=1$ level for 4 different values of the incident coronal 
irradiance on the FAL-C atmospheric model. {\em Bottom:} Same profiles but 
normalized to their maximum.
Note the sensitivity of the relative amplitudes of the blue and red components
of He 10830 \AA\ to the incoming CI.
\label{fig:emission-profiles}}
\end{figure}

The top panels of Fig. \ref{fig:emission-profiles} show the 10830 \AA\ (left)
and $D_3$ (right) emergent profiles at the height of maximum emission (this 
is, 2023 km above $\tau_{500}=1$) for four different values of the CI
(0, 2, 5, and 10 times the nominal EUV spectrum). The tendency is 
clear:
the stronger the CI, the greater the emission. As we shall see below,
this is just a consequence of the increase in the population of the triplet
levels with the CI. 
The lower panels of Fig. \ref{fig:emission-profiles}
represent the same intensity profiles when each of them is normalized to
its maximum emission. The feature that strikes our attention most is the 
strong sensitivity of the relative amplitude of the blue component ($I_B$) of 
the He {\sc i} 10830 \AA\ multiplet to the strength of the incoming CI
(lower left panel of Fig. \ref{fig:emission-profiles}). The 
ratio $\cal R$ $= I_B / I_R$ of the amplitudes of the
blue to red components of the multiplet changes substantially with the
amount of ionizing radiation. This is an extremely interesting feature
since it could provide a useful diagnostic tool for the incoming coronal 
irradiance in off-limb observations of the He {\sc i} triplet. On the 
other hand, $D_3$ shows no observable relative changes among its spectral 
components in response to the ionizing radiaton. We will see further on
that this is just a natural consequence of the D$_3$ line not reaching 
enough optical thickness in these models to show a
differential saturation behavior.

\begin{figure}[t!]
\centering
\includegraphics[width=7cm]{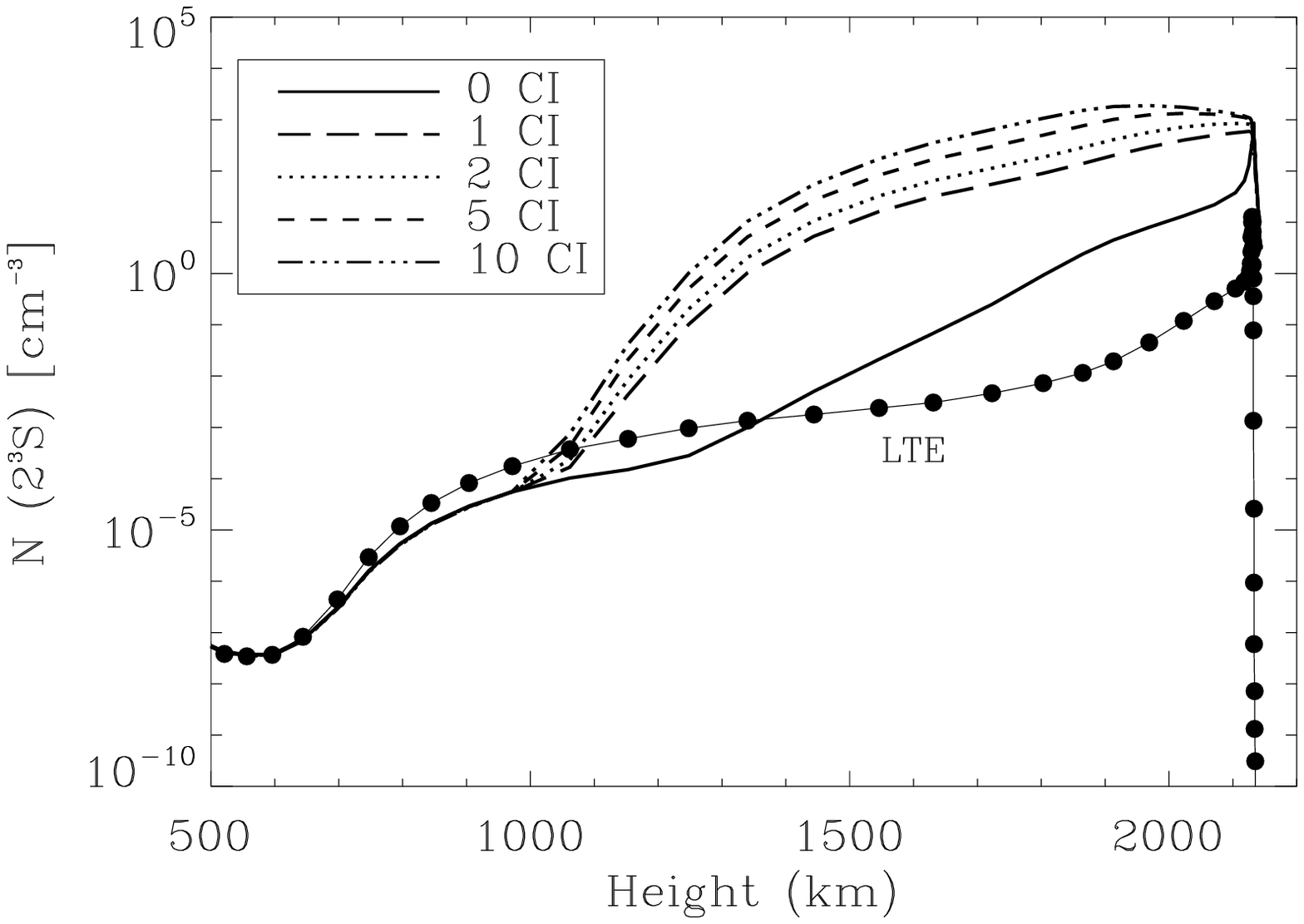}
\includegraphics[width=7cm]{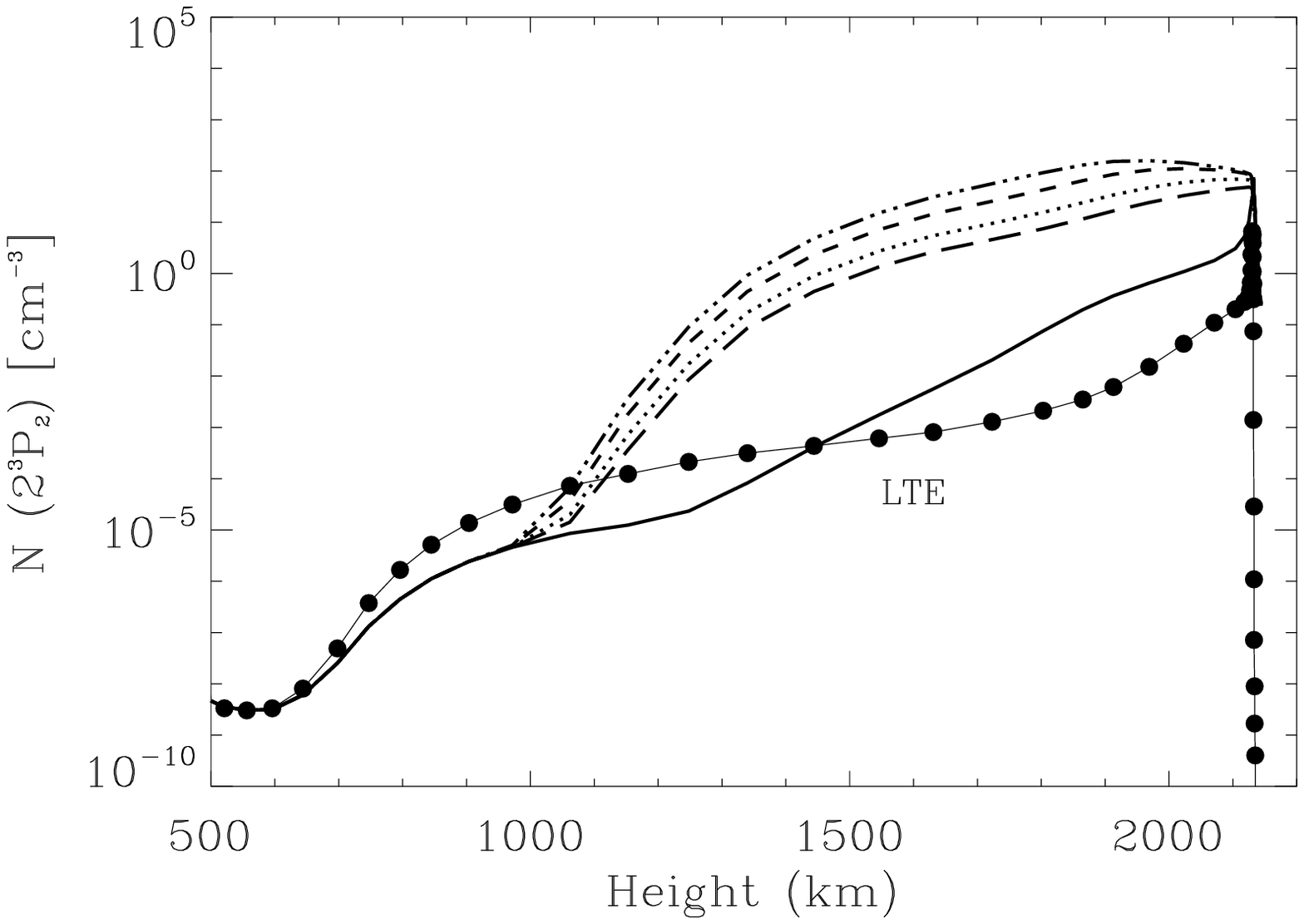}\\
\includegraphics[width=7cm]{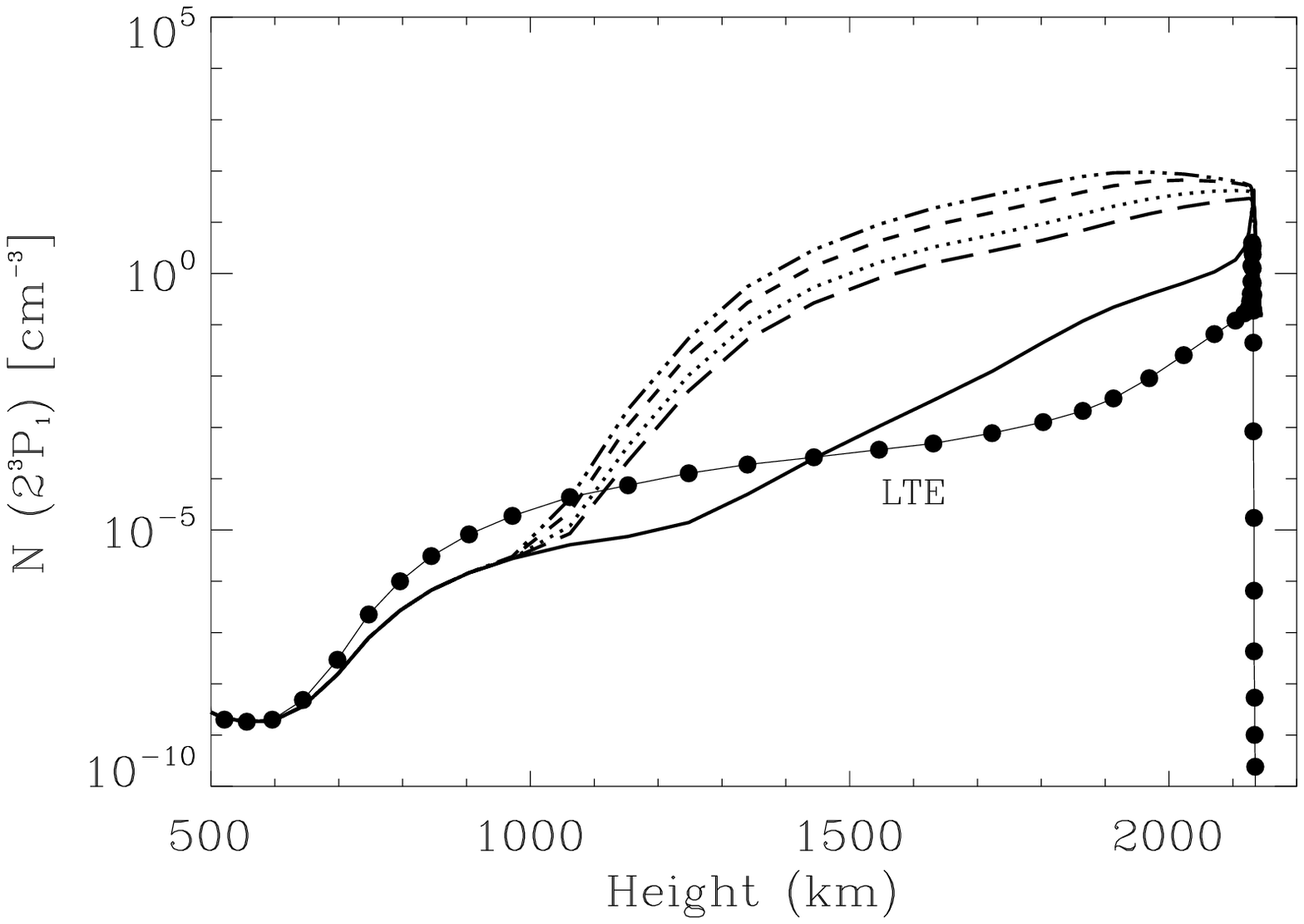}
\includegraphics[width=7cm]{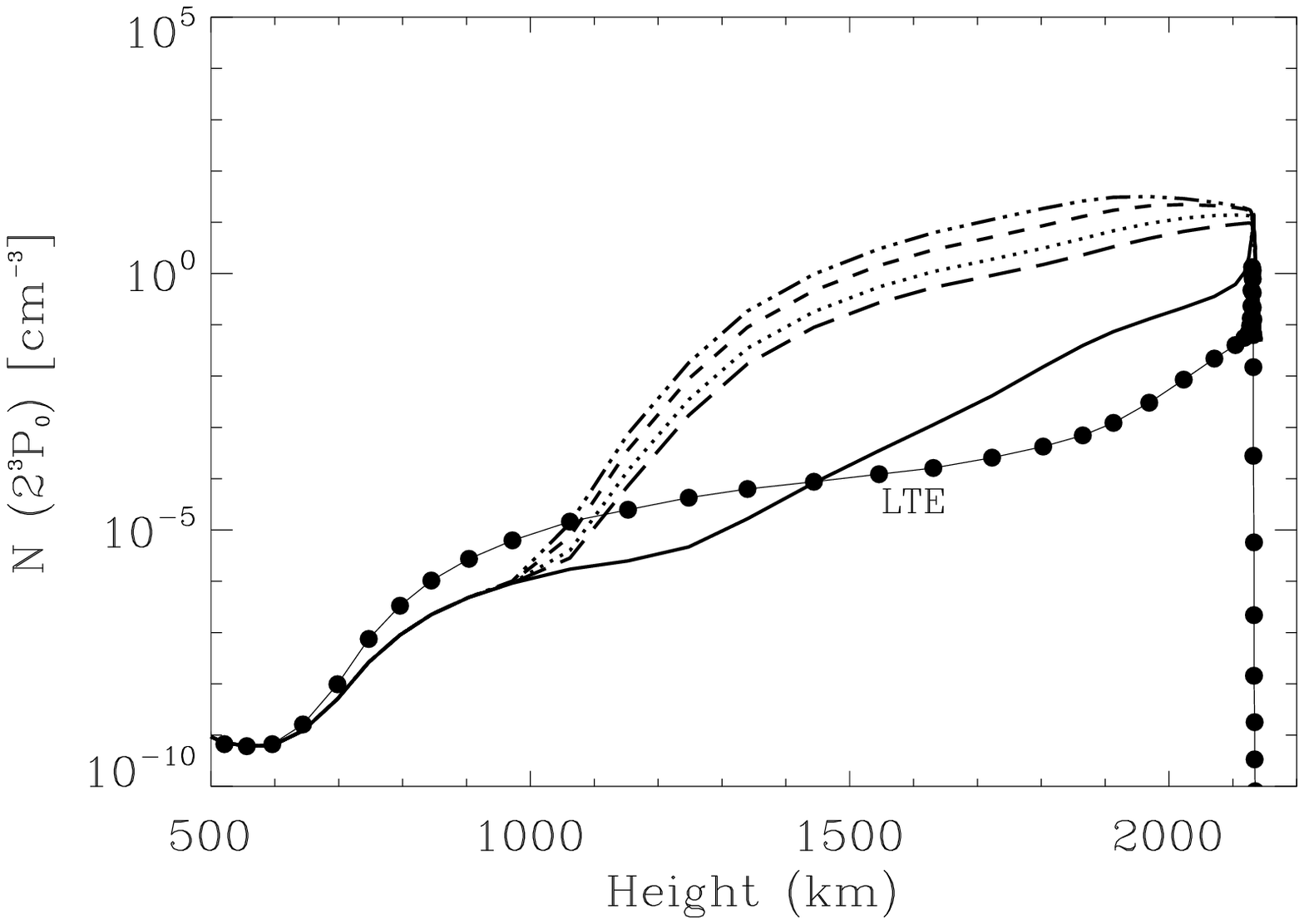}
\protect\caption[ ]{Populations of the energy levels involved in the 
10830 \AA\ transitions (from left to right and top to bottom: $^3S$, $^3P_2$, 
$^3P_1$ and $^3P_0$) as a function of height in the FAL-C model
atmosphere (the reference
level is $\tau_{500}=1$). The populations were computed for 5 different
values of the CI (0, 1, 2, 5 and 10 times its nominal value). Note that
the sensitivity to the ionizing radiation starts at 1000 km above the base
of the photosphere. The solid line with thick dots represents the 
populations in LTE, that, by definition, do not change with the coronal 
irradiance.
\label{fig:populations}}
\end{figure}

In order to understand the changes in the 10830 profiles shown in
Fig. \ref{fig:emission-profiles} let us see how the atomic populations
behave in the absence/presence of the EUV ionizing spectrum.
Fig. \ref{fig:populations} shows the populations of the four $J$ levels 
involved in the 10830 \AA\ transitions ($^3S_1$, $^3P_2$, $^3P_1$ and 
$^3P_0$
from left to right and top to bottom) as a function of height in the
atmosphere, for five different values of the incoming CI (0, 1, 2, 5, and 10 
times the nominal EUV flux).
These plots clearly illustrate how the sole presence of CI dramatically
changes the populations of these levels. Furthermore, the stronger the
CI, the bigger this change. Note, however, that this sensitivity of 
the populations to the CI is only effective starting at a certain height
($\sim 1000$ km) and that it keeps growing until it reaches the sharp
density decrease that characterizes the transition region in the FAL-C
model atmosphere. Below $1000$ km, the extinction of the CI is such that it 
has no effect on the populations. It is interesting to notice that
non-LTE effects play a significant role even in the absence of any coronal
irradiance (compare the solid lines with the LTE results), but the
ensuing overpopulation of the triplet levels above $\sim 1500$ km is
insufficient to produce any significant optical thickness in the 10830 \AA\
lines.

\begin{figure}[t!]
\centering
\includegraphics[width=7cm]{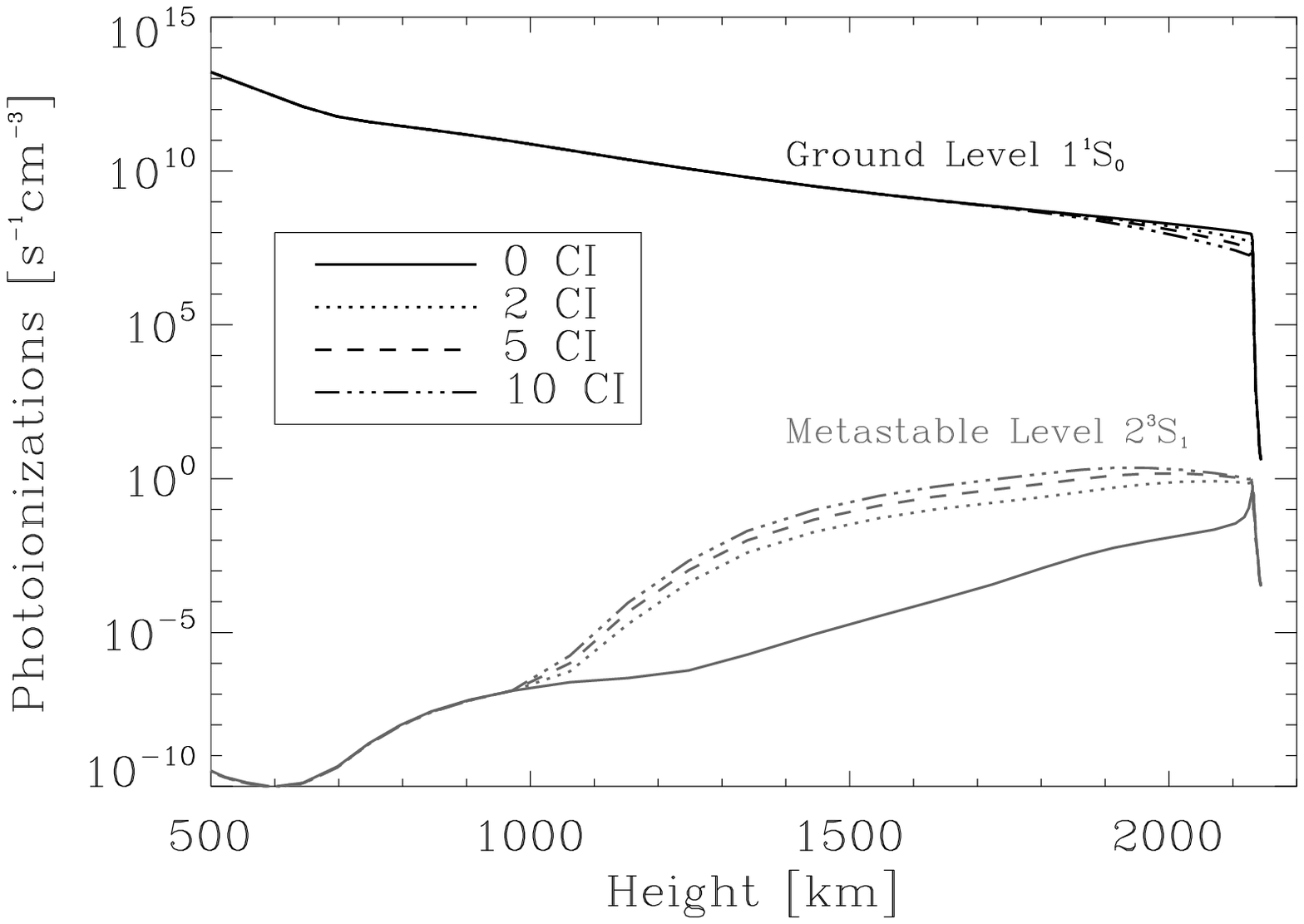}
\includegraphics[width=7cm]{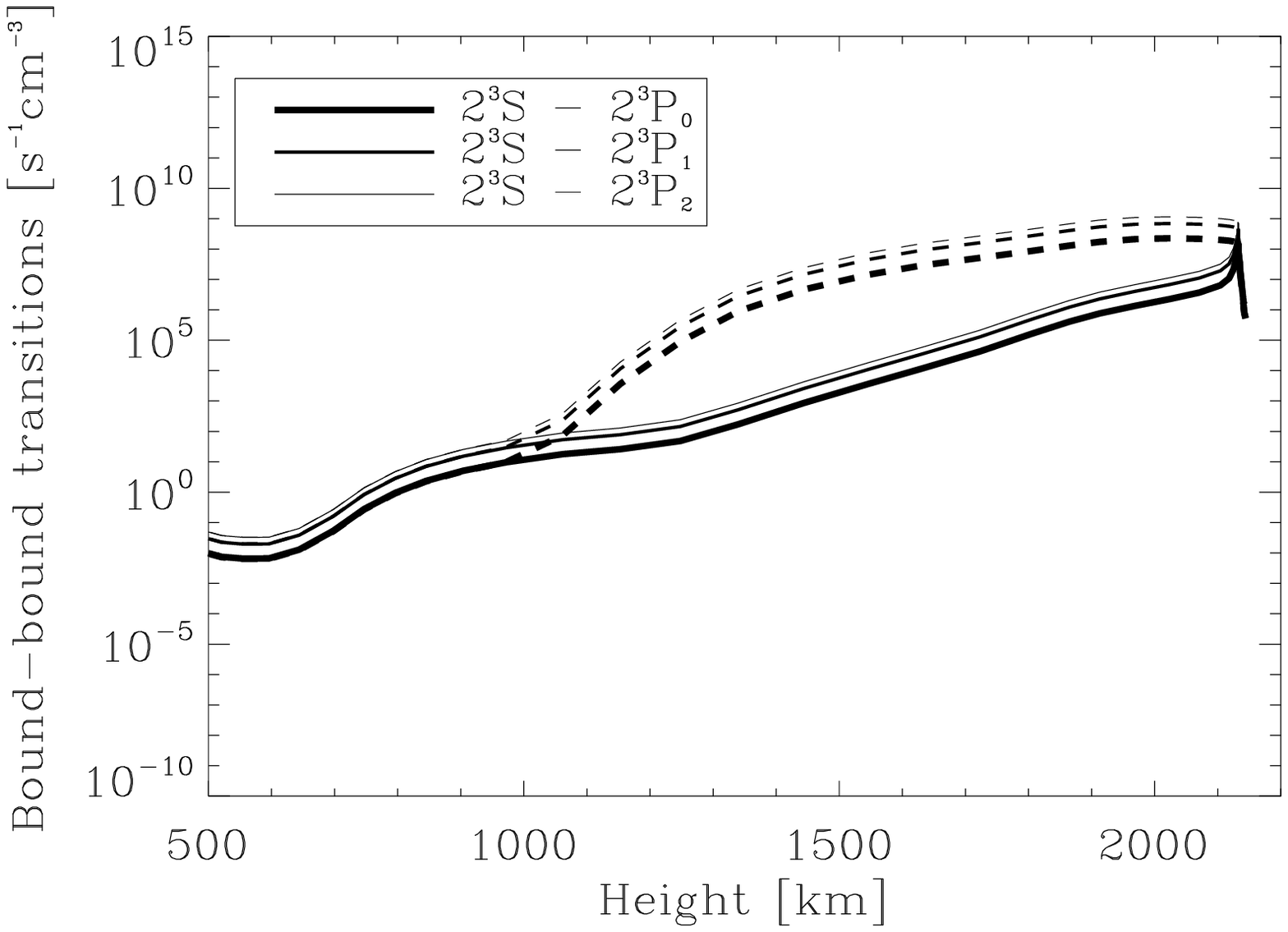}
\protect\caption[ ]{{\em Left}: Number of photoionizations in the FAL-C 
model atmosphere as a function of height. Black and grey show, respectively, 
the photoionizations from the ground and the metastable levels of the 
He {\sc i} atom for 4 different values of the coronal irradiance.
{\em Right}: Number of bound-bound transitions from the metastable level
$2^3S$ to the three upper levels of the 10830 \AA\ multiplet (represented with
different line thickness) as a function of height and for 0 ({\em solid}) and
5 ({\em dashed}) times the nominal coronal irradiance.  
\label{fig:photoionizations}}
\end{figure}

The left panel of Figure \ref{fig:photoionizations} shows the number of
photoionizations per unit volume and time for increasing values of the 
coronal irradiance in the FAL-C model. Notice
that the number of photoionizations that take place from the ground level 
decreases with height, proportionally to the population of this level
- which follows the monotonical decrease of the atmospheric density profile.
On the contrary, the amount of photoionizations taking place from the 
metastable level increases with height, due to the fact that the population 
of this level increases with height between about 600 and 2100 km, even
in the LTE case (see Figure \ref{fig:populations}).
 The effect of the CI is such that it increases
the population of the triplet levels at the expense of the ground level. This
explains why, while the number of photoionizations from the metastable
level grows with increasing CI, the number of photoionizations of
the ground level becomes slightly smaller (due to the decrease in ground level
population).

\noindent It is of some interest to point out that even with 10 times the 
nominal coronal 
value of the EUV flux, the number of photoionizations that take place from
the ground level is ten orders of magnitude larger than that 
taking place from the metastable $^3S_1$ level. Even more interesting
is to keep in mind what the right panel of Figure \ref{fig:photoionizations}
illustrates - this is, that the number of bound-bound 10830 \AA\ transitions
per unit volume and time is way larger than the number of photoionizations 
from the same $^3S_1$ level. The same happens concerning the number of $D_3$
line transitions when compared to the number of photoionizations from their
lower and upper J-levels.

\subsection{Interpretation in terms of the optical thickness of the line formation region}

In a recent investigation, Trujillo Bueno et al. (2005) analyzed
a set of spectropolarimetric observations of quiet Sun chromospheric
spicules at $2''.5$ off the
visible limb in the He {\sc i} 10830 \AA\ multiplet. In their first attempt, 
the authors tried to fit the emergent Stokes profiles ignoring the effects 
of radiative transfer (i.e., using the optically thin approximation). 
However, they found 
that they could not reproduce the emission of the blue component of the 
multiplet under this assumption. In a more realistic approach they were able 
to fit the emission profiles assuming a plane-parallel slab of constant
physical properties and optical thickness $\tau$, that accounts for the 
collective effect of having several spicules interposed along the line of sight.
With a value of $\tau_{R}=3$ (of the optical thickness at the line center 
of the red blended component) the authors were able to reproduce the 
intensity ratio $\cal R$ of the blue and red components of the multiplet, 
proving that radiative transfer effects are non-negligible.

\begin{figure}[t!]
\centering
\includegraphics[width=7cm]{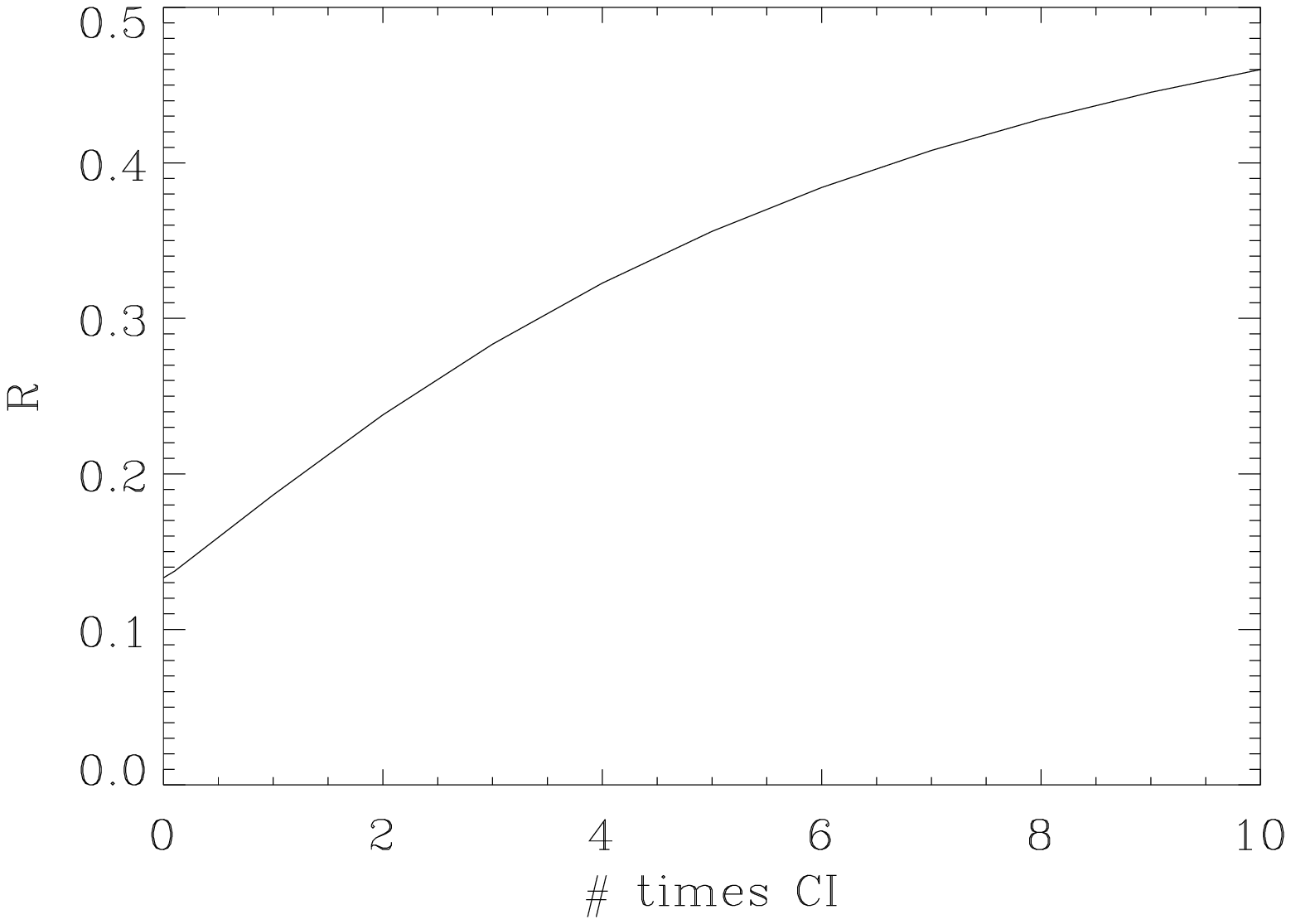}
\includegraphics[width=7cm]{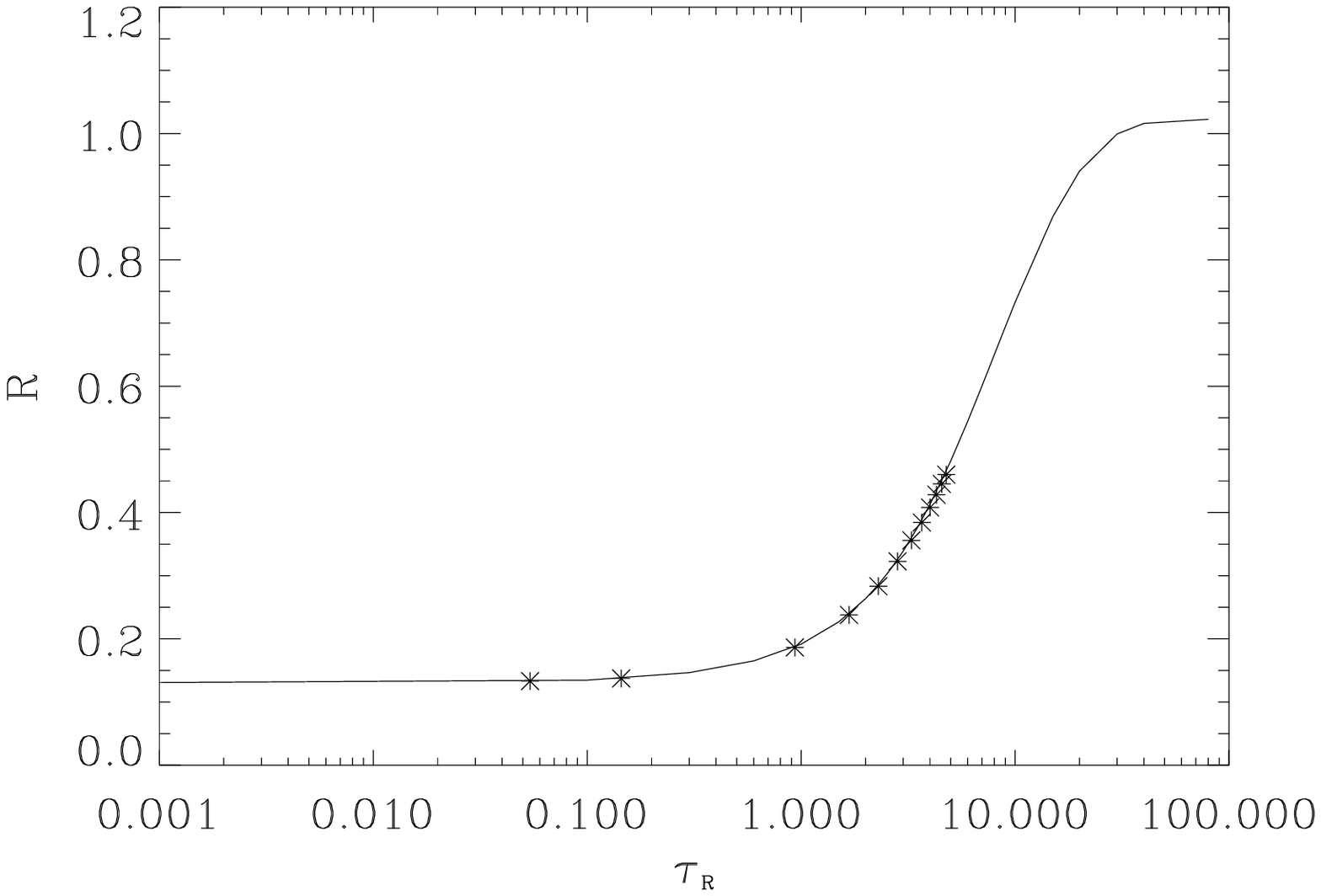}
\protect\caption[ ]{Influence of the coronal irradiance on the optical 
thickness and the ratio $\cal R$ in the FAL-C
model atmosphere. {\em Left}: $\cal R$ as a function of 
the number of times the nominal CI that was applied in the calculations,
for the emergent profiles at 2023 km above $\tau_{500}=1$.
{\em Right}: The solid line represents $\cal R$ as a function of the 
optical depth at the line center of the red component of the multiplet as
predicted by the optically thick slab modeling of Trujillo Bueno et al. (2005).
Asterisks superposed to the solid line are the values obtained from the
profiles simulated in the FAL-C model at the 
aforementioned height.\label{fig:ratio-vs-tau}}
\end{figure}

\noindent The solid line in the right panel of Fig. \ref{fig:ratio-vs-tau} 
illustrates
how the optical thickness of a slab of constant properties modifies the 
emission ratio $\cal R$ of the two components of the 10830 \AA\ multiplet. 
In the optically thin regime ($\tau_{R} < 1$), $\cal R$$=I_B/I_R$
takes a value around $\sim 0.12$, which corresponds to the nominal ratio
of oscillator strengths. For larger values of $\tau_{R}$, this ratio $\cal R$
increases with optical thickness until it reaches a saturation value of
$\sim 1$ for $\tau_{R}\sim 30$.
In this simple modeling, which is similar to that of Trujillo Bueno
et al. (2005), the ratio $\cal R$ is a function of $\tau_{R}$
in a wide range of optical thickness values. But what is the meaning
and physical origin of this parametrization? Qualitatively, the relation
between $\tau$ and the EUV coronal irradiation is easy to
understand. The ionizing radiation modifies the population of the metastable
level $2^3S$, which in time is responsible for the optical thickness of the
transition. But quantitatively, is this $\tau_{R}$ value chosen ad hoc
by Trujillo Bueno et al (2005) to fit the profiles representative of the
physical conditions of the solar corona and of the real integrated optical
thickness along the line of sight?

To address this question we synthesized the He {\sc i} 10830 \AA\ profiles
at a height of 2023 km above the base of the photosphere in the FAL-C
model atmosphere, for different values of the incident coronal irradiance. 
Then, we computed the integrated optical depths along the 
line-of-sight (LOS) at the line center 
wavelength of the red blended component of the multiplet for each case. 
The left panel of Fig. \ref{fig:ratio-vs-tau} relates the measured
ratio $\cal R$ to the amount of inward 
ionizing radiation for a fixed distance to the limb. As expected, the stronger
the CI, the greater the value of $\cal R$, although this relation
is far from being linear and tends to saturate for very large values of
the coronal irradiance.
Asterisks superposed to the right panel of Fig. \ref{fig:ratio-vs-tau} show 
these computed values of $\cal R$ as a function of the integrated optical 
depth along the line-of-sight. The excellent agreement with the 
constant-property slab modeling of Trujillo Bueno et al (2005) evidences that 
the ratio $\cal R$ depends only on the integrated $\tau$ along the LOS, rather
than on the spatial distribution of opacities. 

When $\cal R$ achieves values larger than the ratio of the oscillator 
strengths it means that the transitions are not in the optically thin regime
anymore. In Figure \ref{fig:emission-profiles} the reader can see that, while
the normalized emission intensity profiles of the He {\sc i} 10830 \AA\ 
multiplet show a strong sensitivity to the
incoming CI, He D$_3$, on the other hand, does not. This is a consequence of
the distribution of populations in the triplet system of the He atom. The
lower energy level of the 10830 \AA\ transition is metastable, thus implying 
that its population is orders of magnitude larger than that of the rest 
of the triplet levels (see Figure \ref{fig:populations}). This produces 
a larger opacity in the 10830 \AA\ multiplet that in turn affects the 
ratio $\cal R$. This is an interesting feature because, given the density 
profile of the atmosphere, this ratio should be a reliable tool for 
diagnosing the ionizing coronal irradiation, and viceversa.

\subsection{Change of $\cal R$ with height in various model atmospheres}

\begin{figure}[t!]
\centering
\includegraphics[width=7cm]{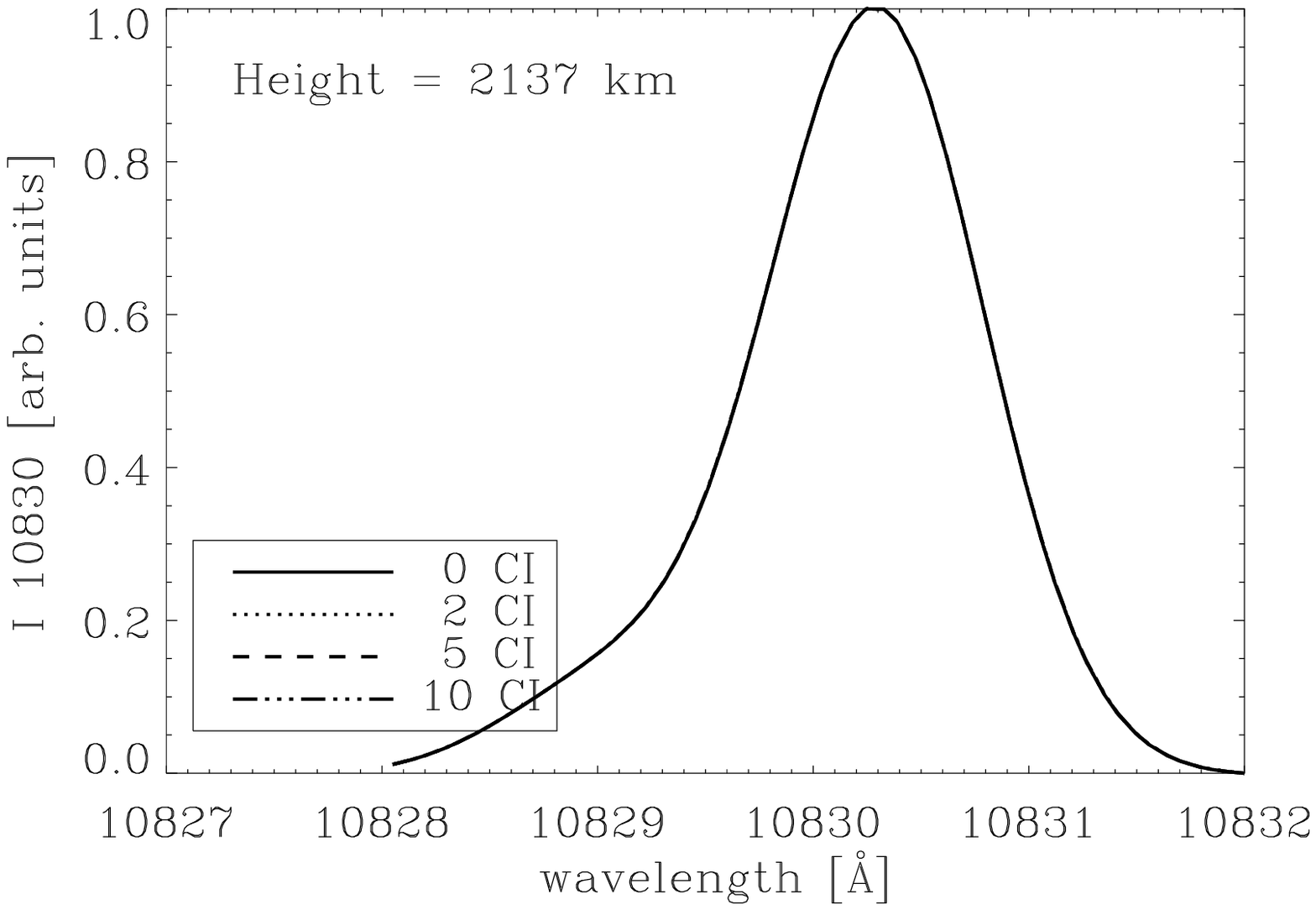}
\includegraphics[width=7cm]{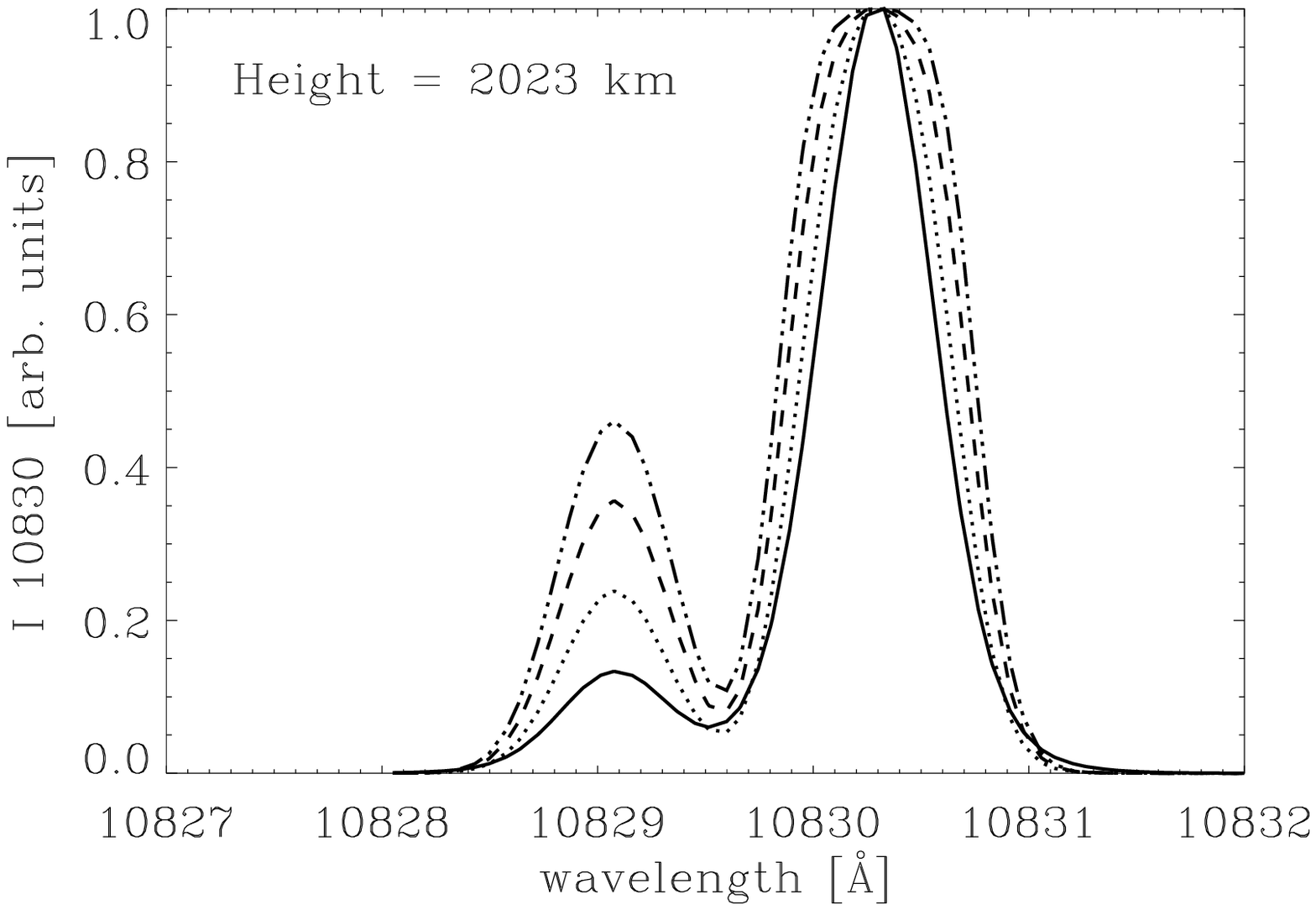}
\includegraphics[width=7cm]{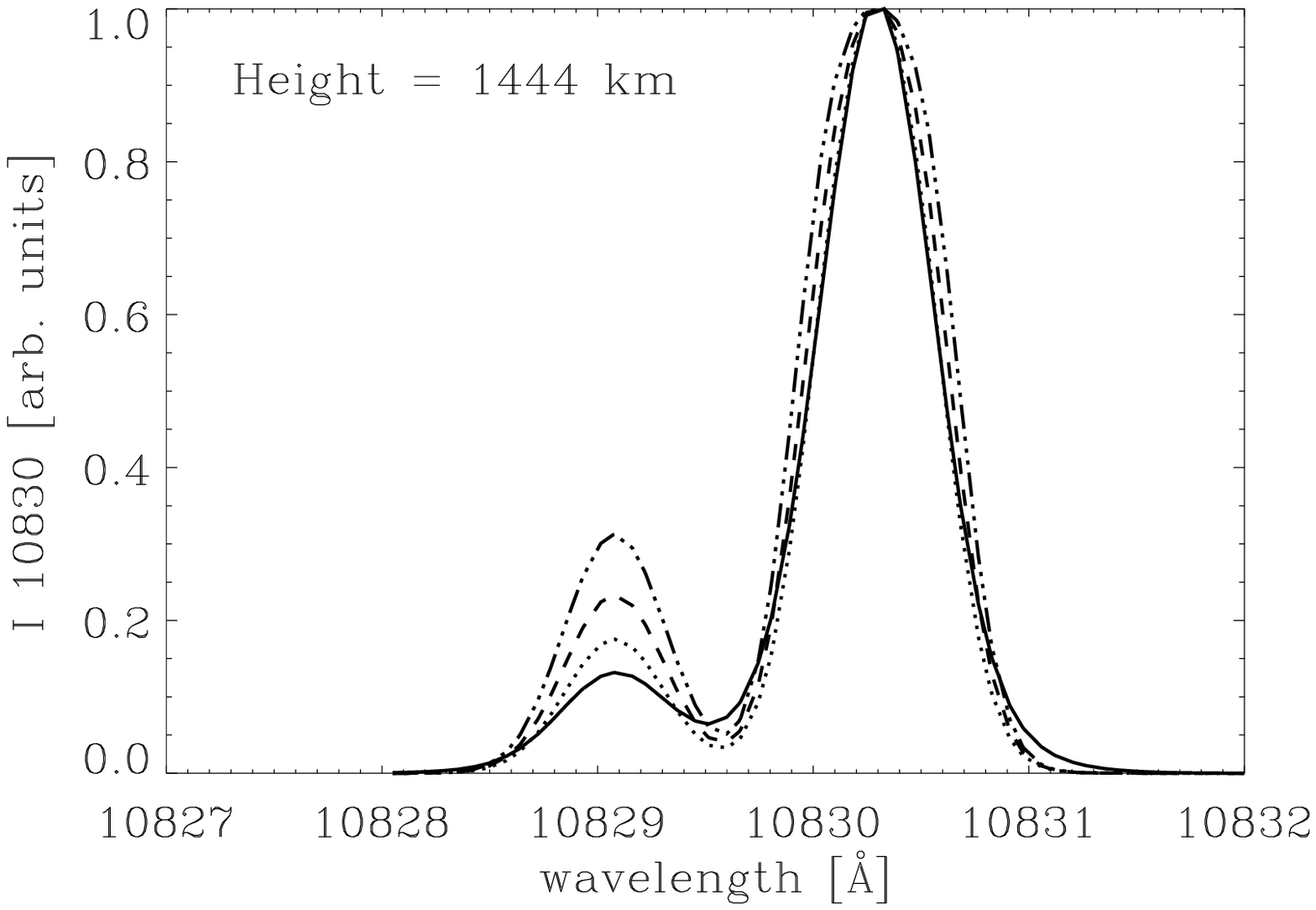}
\includegraphics[width=7cm]{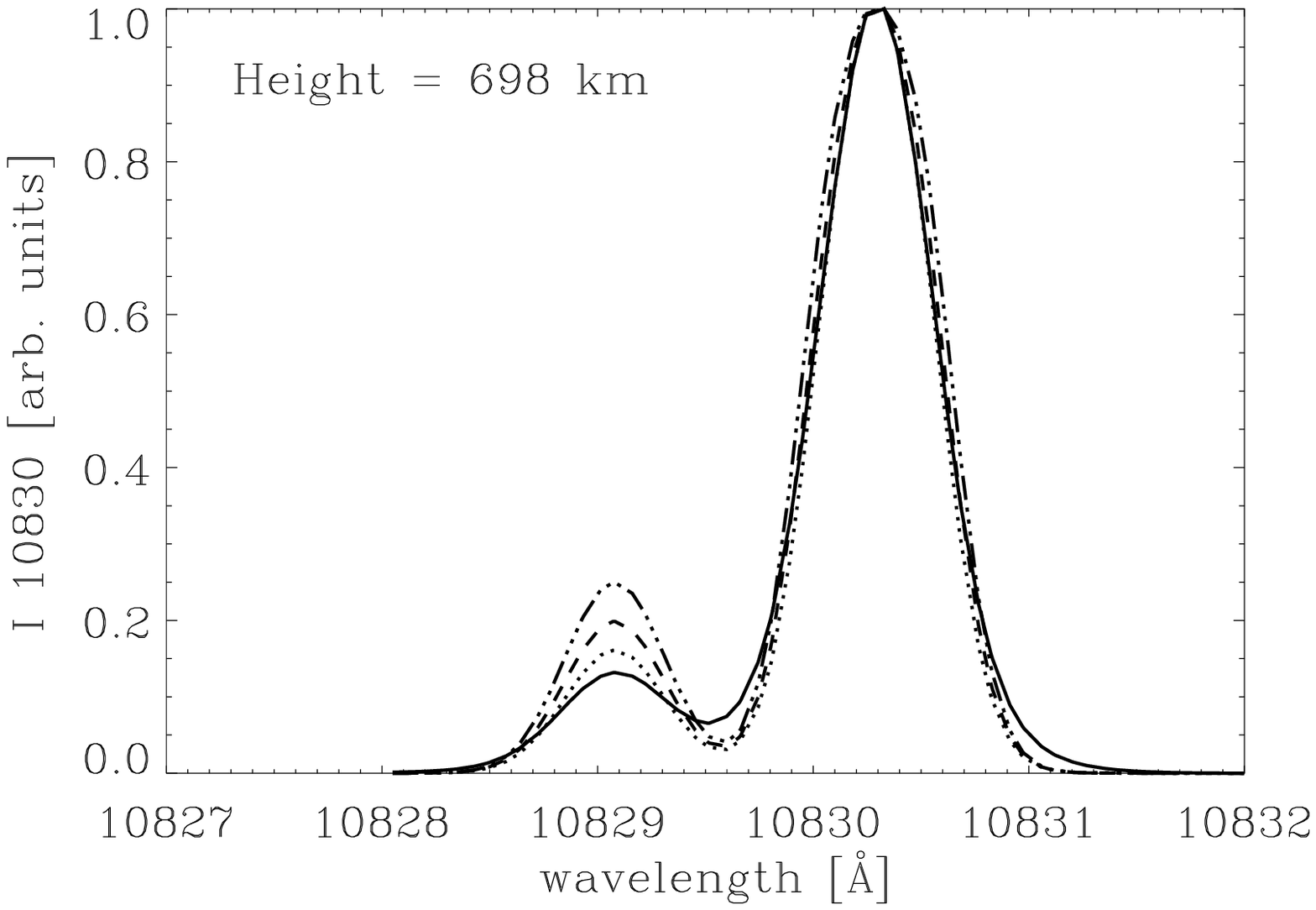}
\protect\caption[ ]{Normalized off-the-limb He {\sc i} 10830 \AA\ intensity 
profiles computed in FAL-C model atmosphere at different distances to the
limb (from left to right and top to bottom, 2137 km, 2023 km, 1444 km and 
698 km above $\tau_{500}=1$).  Each panel shows the intensity spectra for
0, 2, 5, and 10 times the nominal coronal irradiance). 
\label{fig:intensity-vs-distance}}
\end{figure}

The population of the metastable level changes with height, thus, the 
optical thickness in the 10830 \AA\ transitions (and the relative
amplitudes of the blue-to-red components of the multiplet, $\cal R$) should 
behave accordingly.

Fig. \ref{fig:intensity-vs-distance} illustrates the normalized emergent 
intensity profiles in the FAL-C model atmosphere at different 
heights above the limb
(from left to right and top to bottom: 2137 km, 2023 km, 1444 km and 698 km 
above $\tau_{500}=1$). Each panel shows the emission spectra for four values 
of the CI (0, 2, 5 and 10 times its nominal value). This figure proves how
$\cal R$ changes, not only with the amount of CI inciding on the chromosphere,
but also with height in the atmosphere. This is an obvious consequence 
of the dependence of $\cal R$ with the integrated optical depth along the
ray path, and thus with the density.
The top left panel of Fig. \ref{fig:intensity-vs-distance} corresponds to
the emergent profiles from one of the outer-most layers of the FAL-C model. 
The three components of the multiplet appear completely overlapped due to the
large values of the microturbulent velocity ($\sim 8 - 10$ kms$^{-1}$) at
these heights. The widening effect that this contribution produces on the
profiles is larger than the separation between the spectral components.
The closer we get to the limb, the smaller the microturbulence and the
easier it becomes to distinguish the red and blue components of the triplet.
In the FAL-C model, within this range of CIs, the maximum emission (and
ratio $\cal R$) is found at $\sim 2023$ km above $\tau_{500}=1$, although
in general, the stronger the CI, the lower in the atmosphere this would
take place.

\begin{figure}[t!]
\centering
\includegraphics[width=7cm]{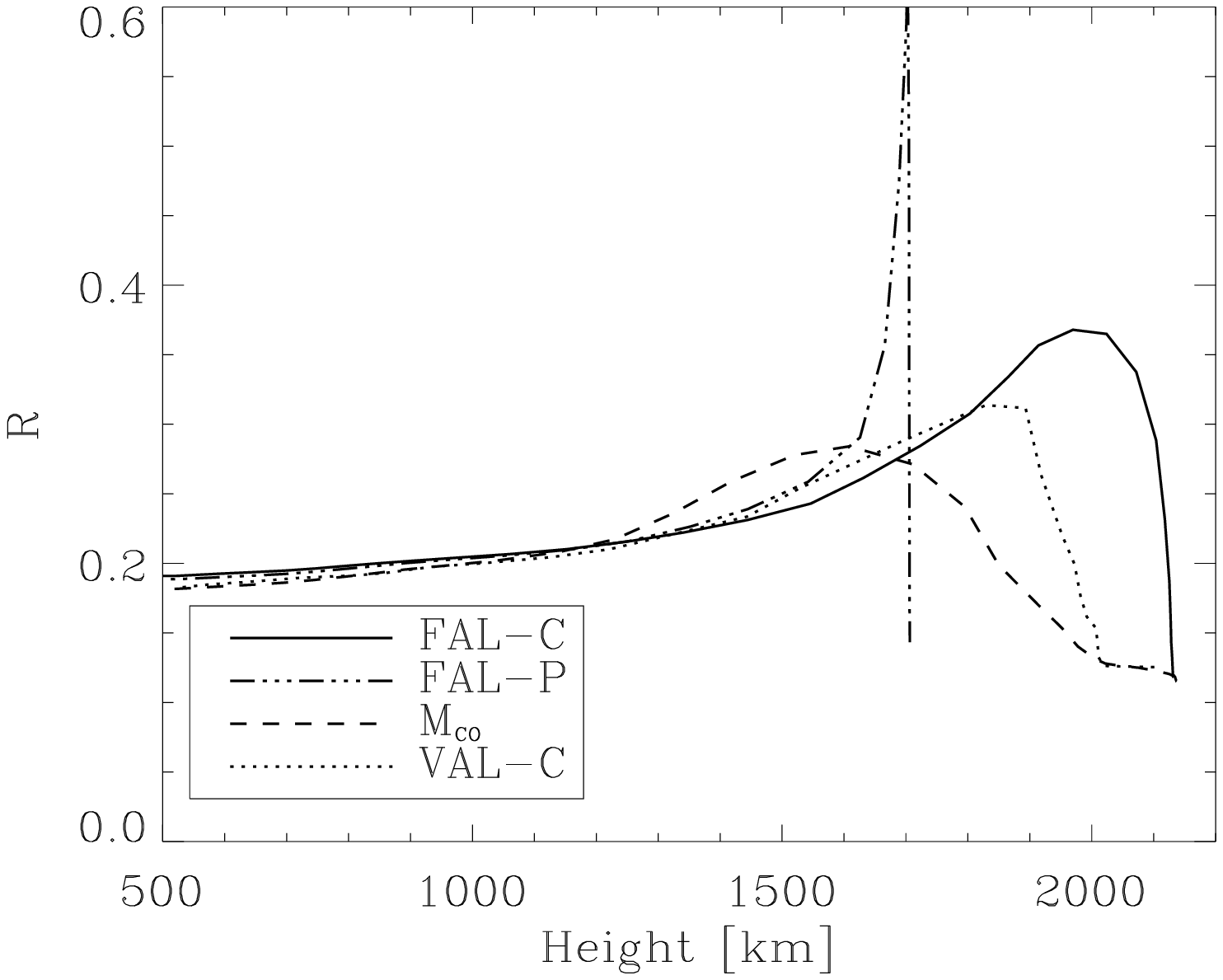}
\includegraphics[width=7cm]{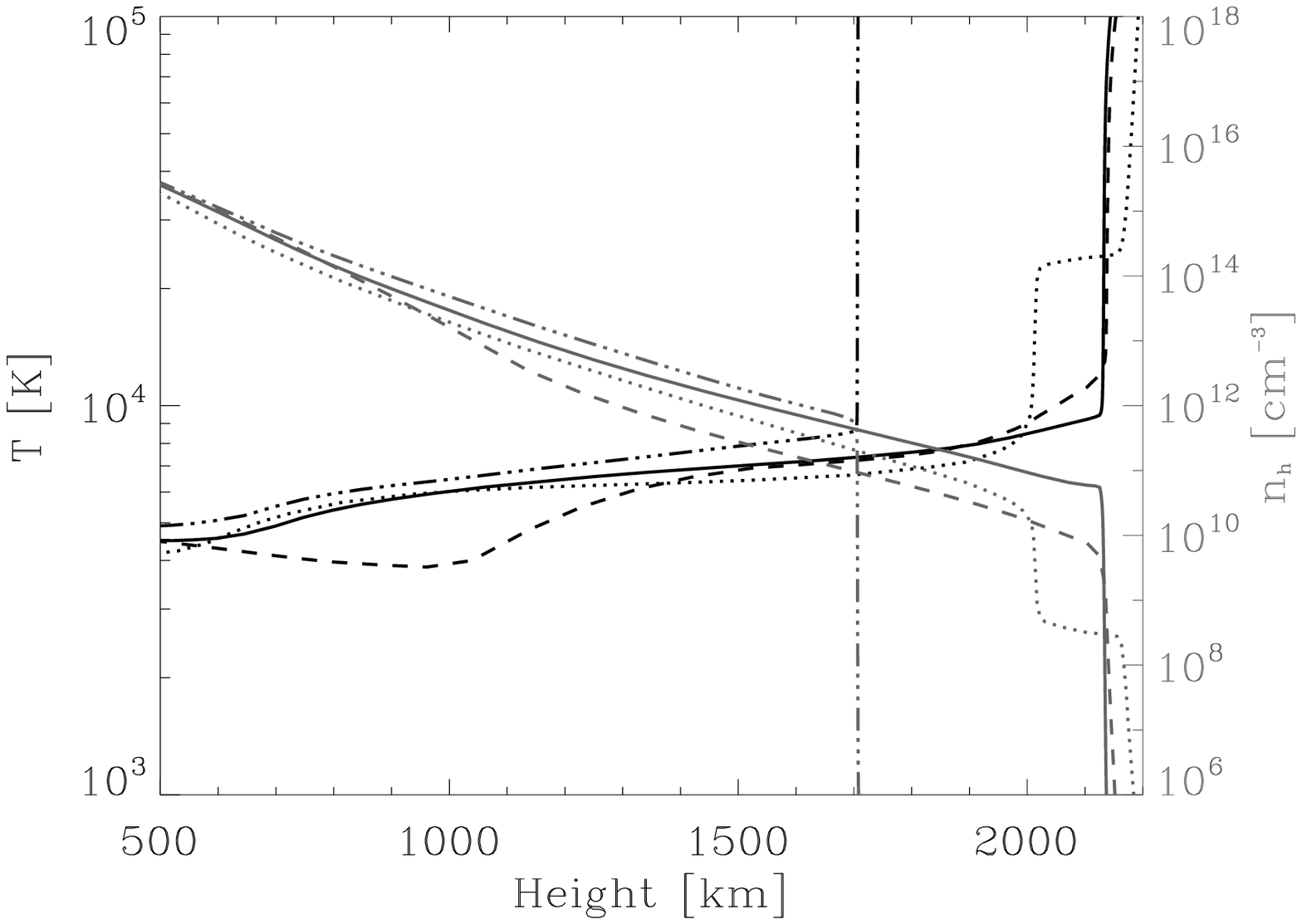}
\protect\caption[ ]{Comparison in 4 model atmospheres: FAL-C, FAL-P, 
VAL-C and the M$_{CO}$ model of Avrett (1995).
{\em Left}: Ratio $\cal R$ of the blue-to-red components of the emergent
He {\sc i} 10830 \AA\ intensity profiles as a function of height in the 
atmosphere when the applied CI takes 5 times its nominal value. 
{\em Right}: Hydrogen number density and temperature profiles as a 
function of height in the four model atmospheres. 
\label{fig:r-with-height}}
\end{figure}

The top left panel of Fig. \ref{fig:r-with-height} shows the ratio $\cal R$ 
as a function of height in four different standard model atmospheres 
(FAL-C, FAL-P, VAL-C and the M$_{CO}$ model of Avrett 1995) when the 
applied CI flux takes five times its nominal value.
The amplitudes of the blue and red components were extracted from a double 
gaussian fitting to the emergent intensity spectra. The higher layers of the
atmosphere, where the overlap between the two components due to the 
microturbulence did not allow an unambiguous determination of $\cal R$, 
were excluded from the calculation (this is, above 1700 km in FAL-P and 
over 2100 km in VAL-C, FAL-C and the M$_{CO}$ model of Avrett 1995).
The very different behavior of the four curves is a consequence of the 
differences between the density profiles and the vertical extent of 
the model atmospheres. 
The right panel of Fig. \ref{fig:r-with-height} illustrates the density 
({\em gray}) and temperature ({\em black}) stratifications in these four 
atmospheres. While FAL-P has a lower transition region (TR) boundary
and a shallower chromosphere, all the other model atmospheres extend much 
higher and
have thicker chromospheres. The lower height of the TR boundary in the 
FAL-P atmosphere combined with the effect of the strong ionizing radiation 
coming downwards from the corona, shifts the region of formation of the He triplet 
towards lower heights in this model. Higher up, the medium becomes 
too rarified to produce enough opacity at the 10830 \AA\ wavelength. 
The density profile of the M$_{CO}$ model of Avrett (1995) in the transition region is not 
as steep as in FAL-C and FAL-P. For this reason, $\cal R$ does not decrease 
as fast in the highest layers of the atmosphere.

\begin{figure}[t!]
\centering
\includegraphics[width=7cm]{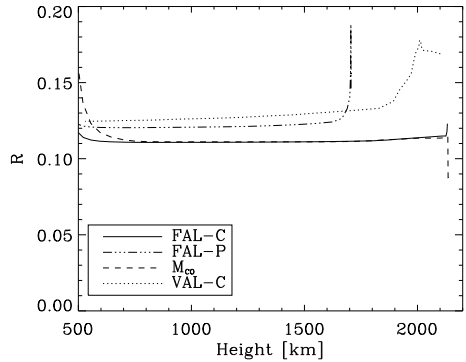}
\protect\caption[ ]{Comparison in 4 model atmospheres (FAL-C, FAL-P,
VAL-C and the M$_{CO}$ model of Avrett 1995) of the ratio $\cal R$ as a 
function of height for the case with
no coronal irradiance inciding on the chromosphere. Note the effect of
the temperature ``plateau'' of the VAL-C model in the region between 
1900 and 2100 km. Also FAL-P shows a very localized peak produced by
its sudden temperature increase at 1700 km above $\tau_{500}=1$.
\label{fig:r-with-height-0ci}}
\end{figure}

\noindent The case for VAL-C is somewhat different. The presence of a 
temperature ``plateau'' at a height of 2100 km that defines a very thick transition region (of a 
couple hundred km) produces an increment of the population of the metastable
level in this layer, even with no help from the ionizing radiation (cf. 
Andretta \& Jones 1997).
This is easy to see in Figure \ref{fig:r-with-height-0ci} where we 
represent the behavior of $\cal R$ as a 
function of height in the four model atmospheres for the case of no coronal 
irradiance inciding on the chromosphere. The temperature plateau and
the thickness of the hot TR of VAL-C make possible a significant number of
collisional excitations that lead to an over-population of 
the triplet system enough to produce a non-negligible optical thickness in 
the 10830 \AA\ transition, but only for off-limb line-of-sights
around 2100 km. Also FAL-P shows a very localized peak
produced by the sudden increase in the temperature stratification at
1700 km above the level $\tau_{500}=1$.

\section{Concluding comment: a diagnostic tool of EUV coronal irradiance}

Figure \ref{fig:r-with-height} shows the change of $\cal R$ with height
in four different model atmospheres. This variation is a very interesting 
feature since it sets a constraint on the thermodynamical structure of
the chromosphere and the amount of ionizing coronal irradiance inciding on it. 
$\cal R$ is an observable than can easily be determined
doing spectroscopic measurements of the He {\sc i} 10830 \AA\ multiplet.
In fact, S\'anchez-Andrade Nu\~no et al. (2007) made use of
this theoretical prediction and have recently presented a series of
observations of off-the-limb He {\sc i} 10830 \AA\ 
spectropolarimetric profiles aimed at the determination of $\cal R$. When
comparing the observations with our radiative transfer modeling, they find 
that the theoretical behavior
of the ratio $\cal R$ agrees qualitatively with the observed one, although 
a quantitative comparison shows poor agreement. 
The density stratification and the limited vertical extent of the 
semiempirical model atmospheres considered in this paper are not adequate 
for spicule modeling. 
The fact that the solar chromosphere is very inhomogeneous on many scales and
that spicules are small-scale intrusions of chromospheric matter into the 
hot corona, make it difficult for 1D models to reproduce off-limb 
observations.

The next step in this study should be to include in our 
RT modeling the spectral line polarization caused by the joint
action of the Hanle and Zeeman effects, as done by Trujillo Bueno \&
Asensio Ramos (2007) within the framework of the above-mentioned slab
model. New spectropolarimetric observations, at different positions above
the solar limb with different levels of ionizing radiation (i.e. active 
regions and coronal holes) should be compared 
to the theoretical calculations to help constrain the atmospheric models
and the physical processes that induce the formation of the He {\sc i}
10830 \AA\ and D$_3$ multiplets.

\acknowledgments

This research was partially funded by the Spanish Ministerio de Educaci\'on 
y Ciencia through project AYA2007-63881.
The National Center for Atmospheric Research (NCAR) is sponsored by the 
National Science Foundation.

\end{document}